\documentclass[a4paper,11pt]{article}

\pdfoutput=1
\usepackage[T1]{fontenc}

\usepackage{xurl}
\usepackage{jcappub}
\usepackage{here}
\usepackage{comment,braket}
\usepackage{epsf}
\usepackage{amsmath}
\usepackage{graphics}
\usepackage{amsfonts}
\usepackage{dblfloatfix}
\usepackage{amssymb}
\usepackage{latexsym}
\usepackage{color}
\usepackage{feynmp}
\usepackage{bm}
\usepackage{cancel}
\usepackage{booktabs}
\usepackage{multirow}

\definecolor{phthaloblue}{rgb}{0.0, 0.06, 0.54}
\hypersetup{
    colorlinks=true,
    linkcolor=phthaloblue,
    citecolor=blue,
    filecolor=blue,
    urlcolor=phthaloblue,
    }

\newcommand{\preprint}[1]{
  \begin{flushright}
    #1
  \end{flushright}
}
    
\title{\boldmath{Interpretation of Nanohertz Gravitational Waves via Cosmic Strings in Post-BBN Matter and Kination Domination}}
\author{Raymond T. Co,}
\author{Siu Cheung Lam,}
\author{Taegyu Lee}
\emailAdd{rco@iu.edu}
\emailAdd{siuclam@iu.edu}
\emailAdd{taeglee@iu.edu}

\affiliation{Physics Department, Indiana University, Bloomington, IN 47405, USA}

\abstract{Recent pulsar timing array (PTA) observations have reported evidence for a common low frequency signal consistent with a stochastic gravitational wave background. In this work, we investigate the cosmic string interpretation of this signal in a non-standard cosmological history between big bang nucleosynthesis and recombination, featuring an early matter-dominated era followed by a kination phase. Previous studies have shown that kination can significantly modify the cosmic-string-induced gravitational wave spectrum, generating characteristic features in the nanohertz range and producing a shape consistent with current PTA data. Building on this, we perform separate Bayesian analyses of the NANOGrav 15-year and EPTA DR2 New+ datasets to quantify the extent to which such a cosmological history is favored by the data. We find substantial evidence for a transient period of matter and kination domination over the standard cosmological history within a cosmic-string interpretation of the gravitational-wave signal, with a Bayes factor $3.63^{+0.22}_{-0.21}$ from the NANOGrav 15-year dataset, while leaving the inferred cosmic-string tension nearly unchanged. This cosmic-string-dominated scenario is particularly relevant if the contribution from supermassive black hole binaries (SMBHBs) is small, as hinted at by astrophysical population simulations. Compared with a scenario where gravitational waves are produced purely by SMBHBs, cosmic-string interpretations with and without an additional contribution from SMBHBs, across both expansion histories and datasets, reveal no significant preference.}

\begin{document}
\preprint{CETUP2026-014}

\maketitle

\flushbottom
    
\section{Introduction}\label{sec:Introduction}
The growing sensitivity of pulsar timing arrays (PTAs) has opened a new observational window onto the nanohertz gravitational wave (GW) regime. Recent results by the NANOGrav 15-year collaboration~\cite{NANOGrav:2023gor}, the European Pulsar Timing Array (EPTA)~\cite{EPTA:2023akd}, the Parkes Pulsar Timing Array (PPTA)~\cite{Zic:2023gta} and the Chinese Pulsar Timing Array (CPTA)~\cite{Chen:2025qqb} have reported evidence for a stochastic GW background (SGWB) in the nanohertz regime. These developments have generated considerable interest since the PTA signal may provide a direct probe not only of astrophysical sources such as supermassive black hole binaries (SMBHBs), but also of relic processes that may have occurred in the early Universe. Possible early-Universe cosmological sources relevant to the PTA band include cosmic strings (CS)~\cite{Vachaspati:1984gt,Caldwell:1991jj}, first-order phase transitions~\cite{Hogan:1986dsh,Kamionkowski:1993fg}, scalar-induced gravitational waves~\cite{Ananda:2006af,Baumann:2007zm}, and gravitational waves associated with non-standard cosmological histories such as kination~\cite{Co:2021lkc,Harigaya:2023pmw}.

In the early Universe, cosmic strings can form during the symmetry breaking phase transition via the Kibble mechanism~\cite{Kibble:1976sj,Kibble:1980mv}. Depending on whether the underlying spontaneously broken symmetry is global or gauged, cosmic strings are broadly classified into global and local strings. In this work, we focus on local cosmic strings. Such symmetry-breaking patterns arise in a variety of particle physics models. A common example is the breaking of an extra Abelian gauge symmetry such as $U(1)_{B-L}$, which frequently appears in models for neutrino mass generation based on the seesaw mechanism~\cite{Minkowski:1977sc,Yanagida:1979as}. Local strings may also emerge in grand unified theories whose symmetry-breaking pattern contains an intermediate $U(1)$ symmetry, as in various $SO(10)$ constructions~\cite{Gell-Mann:1979vob,Kibble:1982ae}. These scenarios make local cosmic strings theoretically and phenomenologically well motivated. Once formed, they follow a scaling law where the fraction of energy density in the network compared to the background is fixed. Moreover, on scales much larger than their core width, local strings are often well approximated by Nambu--Goto strings, so their network evolution leads to predictive cosmological signatures, most notably an SGWB produced by oscillating cosmic string loops~\cite{Vilenkin:1981bx,Vachaspati:1984gt,Turok:1984cn,Burden:1985md}. The resulting SGWB depends not only on string parameters such as the tension, emission spectrum, and intercommutation probability, but also heavily on the background expansion history of the Universe~\cite{Co:2021lkc}. Therefore, the GW spectrum generated by cosmic strings can reveal information not only about string microphysics but also about the history of the cosmological evolution.

Among the simplest and well-motivated departures from the standard thermal history is a kination-dominated (KD) era, during which the energy density of the Universe is dominated by the kinetic energy of a scalar field~\cite{Spokoiny:1993kt,Joyce:1996cp}. A particularly natural realization arises in theories containing a complex scalar field with a global $U(1)$ symmetry and a scalar potential whose non-zero vacuum expectation value spontaneously breaks the $U(1)$ symmetry. In addition, higher-dimensional operators explicitly break the symmetry and, for a large initial radial field value, provide a kick in the angular direction that initiates field rotation in the early Universe.
While the radial mode is still displaced from its vacuum value, the scalar-field energy can drive an early matter-dominated (MD) phase. As the radial mode thermalizes, part of the energy stored in the complex scalar field is dissipated, while the approximately conserved charge prevents the energy from being completely depleted. The motion then approaches circular rotation in field space. Once the radial mode settles to the vacuum, the potential energy becomes negligible. In this regime, the energy density is dominated by rotational kinetic energy, giving rise to a kination era. This kination era corresponds to a phase with an equation of state parameter $w \simeq 1$, where the energy density redshifts as $a^{-6}$ with $a$ the scale factor. Since this component redshifts faster than radiation, the kination phase is temporary and the Universe eventually returns to the radiation-dominated (RD) epoch.

The PTA signal is also interesting from the viewpoint of non-standard expansion histories. Previous work has shown that if an early matter-dominated era is followed by kination, with both transitions occurring between big bang nucleosynthesis (BBN) and recombination, the SGWB from cosmic strings can deviate substantially from the nearly flat spectrum expected in the standard radiation-dominated cosmology~\cite{Co:2021lkc}. In particular, this late matter--kination sequence can modify the resulting spectral structure in the nanohertz range. Motivated by this possibility, in this work we perform a Bayesian analysis of the cosmic string interpretation in the presence of such a late matter-to-kination domination history, and assess whether it provides a better description of current PTA data than the standard cosmology.

The remainder of this paper is organized as follows. In Sec.~\ref{sec:kination}, we review axion rotation that can give rise to the kination cosmology considered in this work, characterized by an early matter-dominated era followed by a kination-dominated era. In Sec.~\ref{sec:CS_GW}, we describe GW production from local cosmic strings and detail the resulting spectrum in the presence of the modified expansion history. In Sec.~\ref{sec:bayesian}, we outline the Bayesian analysis framework and describe the use of Bayes factors for model comparison. In Sec.~\ref{sec:Analyses}, we present the constraints imposed by BBN and cosmic microwave background (CMB) observations. The results of the PTA analyses are presented and discussed in Sec.~\ref{sec:results}, followed by our conclusions in Sec.~\ref{sec:conclusion}.

\section{Axion rotation and kination}
\label{sec:kination}
A simple and well-motivated origin of a KD era is provided by the rotation of a complex scalar field that spontaneously breaks an approximate global $U(1)$ symmetry~\cite{Co:2019wyp}.
We write the complex scalar field as
\begin{equation}
P(x)=\frac{1}{\sqrt{2}}\,S(x)e^{i\theta(x)},
\end{equation}
where $S$ denotes the radial mode and $\theta$ the angular mode. In the vacuum, the radial field settles to its vacuum expectation value, $f_a$, so that the canonically normalized axion field is
\begin{equation}
\phi(x)=f_a\,\theta(x).
\end{equation}
The dynamics are described by
\begin{equation}
\mathcal{L} = |\partial P|^2 - V(P) = \frac{1}{2}(\partial S)^2 + \frac{1}{2}S^2(\partial\theta)^2 - V_0(S) - V_{\cancel{U(1)}}(P),
\end{equation}
where $V_0(S)$ is the $U(1)$-symmetric part of the potential, while $V_{\cancel{U(1)}}(P)$ denotes terms that explicitly break the $U(1)$ symmetry. An example is a higher-dimensional operator of the form
\begin{equation}
V_{\cancel{U(1)}}(P) = \left( \frac{P^n}{M^{n-4}} + {\rm h.c.} \right), \qquad n>4,
\end{equation}
with $M$ the cutoff scale of the theory. These terms are expected to exist when the $U(1)$ symmetry is an accidental symmetry~\cite{Holman:1992us,Barr:1992qq,Kamionkowski:1992mf,Dine:1992vx} and is broken by quantum gravity~\cite{Giddings:1988cx,Coleman:1988tj,Gilbert:1989nq,Harlow:2018jwu,Harlow:2018tng}. In the early Universe, the scalar field can acquire a large field value by a negative Hubble-induced mass term during inflation. When the radial field value is large, this operator exerts a torque in the angular direction and sets the field into rotational motion in field space, in close analogy with Affleck--Dine dynamics~\cite{Affleck:1984fy}. 

Once the explicit breaking term becomes negligible due to the redshift of the radial field amplitude, the angular motion is associated with a conserved $U(1)$ charge. For a homogeneous condensate, the corresponding charge density is
\begin{equation}
n_\theta \equiv i\dot P^\ast P - i\dot P P^\ast = S^2 \dot\theta,
\end{equation}
and it redshifts simply as $1/a^3$. Equivalently, in the absence of significant entropy production, the yield
\begin{equation}
Y_\theta \equiv \frac{n_\theta}{s}
\end{equation}
remains approximately constant, where $s$ is the entropy density of the Universe. This conserved charge is the key quantity that controls subsequent cosmological evolution. The important point is that the same rotating condensate can behave first as matter and later as kination, depending on whether the rotation occurs at large radius or close to the vacuum expectation value.

\subsection{Matter-like evolution}

The first stage arises while the radial field is still far from its vacuum value. In well-motivated scenarios such as supersymmetric theories~\cite{Moxhay:1984am,Co:2020jtv}, the radial potential can remain nearly quadratic over a wide range of large field values despite the presence of a nonzero vacuum expectation value. At this stage, we therefore approximate
\begin{equation}
V_0(S)\simeq \frac{1}{2}m_S^2 S^2,
\qquad S\gg f_a.
\end{equation}
During this stage, the orbit is well described by rotational motion with a redshifting amplitude. The angular velocity is determined by the radial equation of motion, which gives
\begin{equation}
\dot{\theta}^2 \simeq \frac{V_0'(S)}{S}.
\end{equation}
For a quadratic potential, $V_0'(S)=m_S^2 S$, and therefore
\begin{equation}
\dot\theta \simeq m_S.
\end{equation}
Since the conserved charge is $n_\theta = S^2 \dot\theta$ and $n_\theta \propto a^{-3}$, the orbit radius decreases as
\begin{equation}
S^2 \propto a^{-3}.
\end{equation}
The energy stored in the rotating condensate is then dominated by the rotational and quadratic-potential pieces, both of which scale in the same way. As a result, the energy density of the rotation follows
\begin{equation}
\rho_{\rm rot} \propto a^{-3}.
\end{equation}
Thus, while the field is rotating at large radius, the condensate behaves effectively as matter. Importantly, this matter scaling is not caused by oscillations around the minimum in the usual case, but by coherent rotation at the large field radius. 

Although radial oscillations are inevitably triggered together with the initial kick, the radial motion can later be depleted during the evolution. If the radial mode interacts with the thermal bath, it can dissipate the energy density stored in the radial mode. This can be done by perturbative decays of the radial mode or by scattering of the radial mode with gauge bosons or with fermions in the thermal bath~\cite{Co:2019wyp}.
Thermalization becomes efficient when $\Gamma_{\rm th} > H$. However, if the interaction responsible for this dissipation does not break the $U(1)$ symmetry, it dissipates energy without affecting the associated charge, i.e., rotation in the angular direction. Thermalization therefore does not remove the entire energy stored in the complex scalar field. Rather, it removes the energy associated with radial motion, rendering the rotational motion circular.

\subsection{Kination-like evolution}

As the Universe expands further, the charge density continues to dilute and the radius of the orbit shrinks. Eventually the radial field approaches its vacuum value. At that point the motion is no longer a rotation at a large radius, but rather a motion along the bottom of the wine-bottle potential with nearly fixed radius.
The conserved charge then implies
\begin{equation}
n_\theta = f_a^2 \dot\theta,
\end{equation}
so that the angular velocity decreases as $\dot\theta \propto a^{-3}$.
The complex field energy density is now dominated by the angular kinetic term,
\begin{equation}
\rho_{\rm rot}
\simeq
\frac{1}{2} f_a^2 \dot\theta^2
=
\frac{n_\theta^2}{2f_a^2}
\propto
a^{-6},
\end{equation}
which is the scaling of kination and corresponds to an equation of state $w = 1$. We assume that the axion is sufficiently light that axion rotation continues past $T_{\rm KR}$.

The essential picture is therefore very simple. At large radius, the energy per unit charge is set by the radial mass scale, and the condensate behaves like matter. Once the radius reaches $f_a$, the same conserved charge is stored in pure angular motion, and the energy density begins to redshift much faster, scaling as $a^{-6}$.

\subsection{Characteristic temperatures}

\begin{figure*}[htbp]
    \centering
    \includegraphics[width=\textwidth]{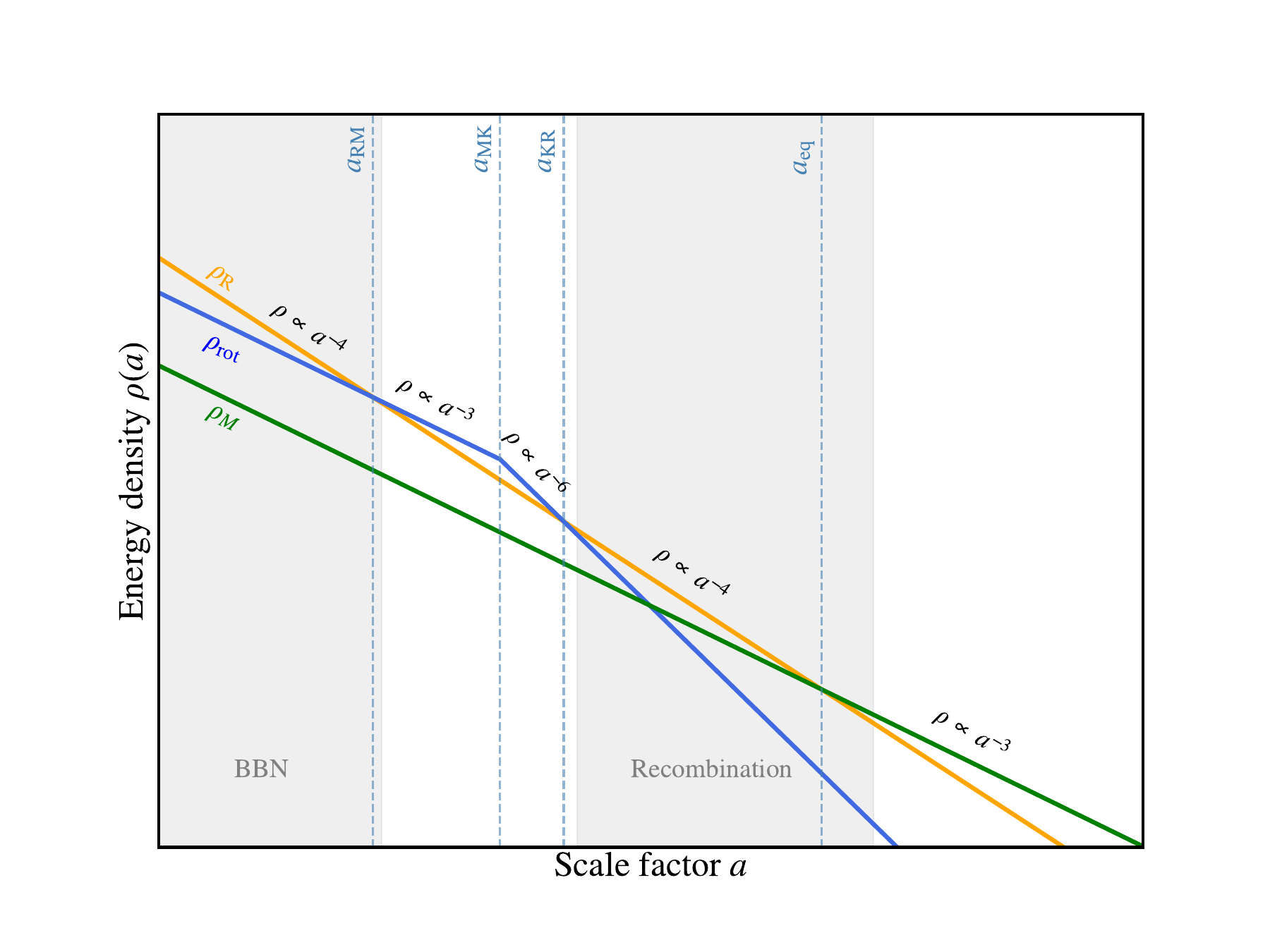}
    \caption{Schematic representation of the cosmological evolution of the different energy components with scale factor \textit{a} in the presence of a rotating field: ${\rm RD}\rightarrow{\rm MD}\rightarrow{\rm KD}\rightarrow{\rm RD}\rightarrow{\rm MD}$. Radiation, matter and scalar field energy densities are denoted by $\rho_R$, $\rho_M$ and $\rho_{\rm rot}$ and are represented by orange, green and blue solid lines, respectively. The vertical light blue dashed lines indicate, from left to right, the RM, MK, and KR transitions, occurring at $a_{\rm RM}$, $a_{\rm MK}$ and $a_{\rm KR}$, followed by standard matter-radiation equality at $a_{\rm eq}$, respectively. In the scenario considered here, the intermediate matter-dominated era arises from scalar field rotation at large radius. As the radial field settles toward its vacuum value, the angular kinetic energy dominates, driving a period of kination phase, after which the radiation component eventually overtakes the rapidly redshifting scalar field energy density, bringing the Universe back to radiation domination. Finally, after matter-radiation equality, standard matter domination ensues until late-time dark energy domination (not shown).}
    \label{fig:energy_density}
\end{figure*}

In this work, both MD and KD are essential in producing the characteristic cosmic string GW spectrum that is relevant for PTA observations. For this cosmological sequence to occur between BBN and recombination, the radial mode should thermalize before the energy density stored in the rotating condensate dominates over the radiation energy density. If this criterion is satisfied, the Universe enters an early MD era and transitions to a KD era. As a result, the cosmological history naturally proceeds from being RD to MD, to KD, and subsequently to RD. This cosmology is illustrated schematically in Fig.~\ref{fig:energy_density}. Importantly, neither transition requires the dominant component to decay into radiation. In particular, unlike the usual early matter domination that ends by reheating, the onset of kination need not be accompanied by entropy production. The eventual transition from MD to KD and then to RD occurs automatically through cosmic redshift.

It is useful to describe this history in terms of the approximately conserved yield $Y_\theta = n_\theta/s$. Below, we denote by $T_{\rm RM}$, $T_{\rm MK}$, and $T_{\rm KR}$ the temperatures at the time of the radiation-to-matter, matter-to-kination, and kination-to-radiation transitions, respectively. During the matter-dominated stage, the rotational energy density is of order
\begin{equation}
\rho_{\rm rot}^{\rm (MD)} \simeq m_S n_\theta = m_S Y_\theta s,
\end{equation}
where $s=2\pi^2g_{\ast s}T^3/45$ is the entropy density.
Equating the condensate energy density with the radiation energy density, $\rho_R=\pi^2g_\ast T^4/30$, gives the temperature at which the rotating condensate begins to dominate,
\begin{equation}
T_{\rm RM}\simeq \frac{4}{3}\,m_S Y_\theta,
\end{equation}
where we have taken $g_{\ast s}\simeq g_\ast$.

The transition from matter domination to kination domination occurs when the orbit radius shrinks to the vacuum value. Using the matter scaling relation between charge and radius one finds
\begin{equation}
T_{\rm MK}
\simeq
\left(
\frac{45}{2\pi^2 g_\ast(T_{\rm MK})}
\frac{m_S f_a^2}{Y_\theta}
\right)^{1/3}.
\end{equation}
Finally, kination ends when the $a^{-6}$ component falls below radiation, namely when
\begin{equation}
\frac{1}{2}f_a^2\dot\theta^2 = \rho_R.
\end{equation}
This gives
\begin{equation}
T_{\rm KR}
\simeq
\frac{3\sqrt{15}}{2\pi\sqrt{g_\ast(T_{\rm KR})}}
\frac{f_a}{Y_\theta}.
\end{equation}
It is worth noting that these transition temperatures are not independent. Up to the effective relativistic degrees of freedom for energy and entropy, $g_\ast$ and $g_{\ast s}$, they satisfy 
\begin{equation}
    T_{\rm MK}^3 \simeq T_{\rm RM} T_{\rm KR}^2.
    \label{MKrelation}
\end{equation}
A genuine KD era requires the matter-to-kination transition to occur before radiation takes over, which can be expressed as
\begin{equation}
T_{\rm MK}\gtrsim T_{\rm KR}.
\end{equation}

The scenario described above is especially relevant for the cosmic string interpretation studied in this work. The stochastic gravitational wave background emitted by a cosmic string network is sensitive not only to the microscopic string parameters, but also to the expansion history of the Universe. Consequently, a sequence such as ${\rm RD}\rightarrow{\rm MD}\rightarrow{\rm KD}\rightarrow{\rm RD}$ can imprint characteristic breaks and changes of slope in the gravitational wave spectrum, potentially shifting the relevant spectral features into the PTA band.

\section{Cosmic string gravitational waves}
\label{sec:CS_GW}

\subsection{Local cosmic strings and their gravitational wave background}
Local cosmic strings can form when a $U(1)$ gauge symmetry is spontaneously broken~\cite{Kibble:1976sj}. In local strings, unlike global strings, the would-be Goldstone mode is eaten by the gauge field, and the resulting defect is therefore a local string. Once they are formed, they quickly reach a scaling regime consisting of two components: $\mathcal{O}(1)$ infinite (long) strings per Hubble volume, whose radius of curvature is of order the Hubble length, $H^{-1}$, together with a population of sub-horizon-sized cosmic-string loops~\cite{Vachaspati:1984dz,Bennett:1987vf}. Infinite strings intercommute with each other and their energy density is continuously transferred to loops. These loops subsequently shrink by emitting gravitational waves, and the present-day stochastic background is obtained by summing the loop emission over the entire cosmic history. The gravitational wave spectrum is calculated as
\begin{equation}\label{eq:GW_spectrum_CS}
\Omega_{\rm GW}(f) \equiv \frac{1}{\rho_c}\frac{d\rho_{\rm GW}}{d\ln f} = \frac{8\pi f}{3H_0^2} \sum_{m=1}^{\infty} G\mu^2 P_m\, C_m(f),
\end{equation}
where $\rho_{\rm GW}$ and $f$ are the GW energy density and frequency, $\mu$ is the string tension, $H_0$ is the Hubble constant today, and $P_m$ denotes the fraction of the total loop power emitted into the $m$-th harmonic,
\begin{equation}
   P_m=\frac{\Gamma}{\zeta(q)}\,m^{-q}, \qquad \sum_{m=1}^{\infty} P_m=\Gamma. 
\end{equation}
Here $\Gamma\simeq 50$ is the total gravitational wave emission efficiency~\cite{Blanco-Pillado:2017oxo}, while $q=4/3$, $5/3$, or $2$ for cusps, kinks, or kink-kink collision dominated emission, respectively. The factor $C_m(f)$ encodes the loop number density and the cosmological redshift,
\begin{equation}
C_m(f) = \int_{t_{\rm scl}}^{t_0} dt'\, \left(\frac{a(t')}{a_0}\right)^3 \frac{dn}{df'}(t',f'),
\end{equation}
where $t_{\rm scl}$ is the time at which the long string network enters the scaling regime, $a_0\equiv a(t_0)$ is the present day scale factor, and $t'$ is the emission time. The loop number density per unit emitted frequency is
\begin{equation}
\frac{dn}{df'}(t',f')
= \left( \frac{F\,C_{\rm eff}(t_k)}{\alpha\,t_k^4} \right) \left( \frac{a(t_k)}{a(t')} \right)^3 \left( \frac{1}{\alpha+\Gamma G\mu} \right) \left( \frac{2m}{f'^2} \right),
\end{equation}
with $F\simeq 0.1$, which characterizes the fraction of the long string energy transferred into loops of initial length $l_k=\alpha t_k$. For a loop formed at time $t_k$, the length at emission time $t'$ is
\begin{equation}
l(t',t_k)=\alpha t_k-\Gamma G\mu\,(t'-t_k),
\end{equation}
so that the emitted frequency in the $m$-th harmonic is
\begin{equation}
f'=\frac{2m}{l(t',t_k)}, \qquad f=\frac{a(t')}{a_0}f'.
\label{eq:freqharmonic}
\end{equation}
Combining these relations gives the loop formation time relevant for a present day observed frequency $f$,
\begin{equation}
t_k(t',f) = \frac{\displaystyle \frac{2m}{f}\frac{a(t')}{a_0}+\Gamma G\mu\, t'}{\alpha+\Gamma G\mu}.
\end{equation}
The loop formation efficiency depends on the background equation of state at $t_k$, and is commonly approximated by~\cite{Cui:2017ufi,Blasi:2020wpy}
\begin{equation}
C_{\rm eff}(t_k)\simeq
\begin{cases}
5.4, & \text{RD},\\
0.39, & \text{MD},\\ 
30, & \text{KD}.
\end{cases}
\end{equation}
This makes explicit that $\Omega_{\rm GW}(f)$ is determined jointly by the loop formation history, the loop decay law, and the background expansion history.

\subsection{GW spectrum in standard cosmology}

Before turning to non-standard expansion histories, it is useful to recall the benchmark spectral behavior expected in the standard cosmological background. In this case, the observed spectrum can be understood analytically by tracking the epoch in which loops are formed and the epoch in which they predominantly emit GW.

An analytic estimate of the spectral shape is obtained by considering loops formed in an era with an equation of state parameter $w_1$ and emitting GW in an era with $w_2$. For the fundamental mode with $m=1$, one finds~\cite{Cui:2018rwi}
\begin{equation}
\Omega_{\rm GW}^{(1)}(f)\propto f^{\lambda},
\end{equation}
with
\begin{equation}
\lambda=
\begin{cases}
\dfrac{w_1(6w_2+4)-2}{(w_1+1)(3w_2+1)}, & t_{\Gamma}<t_0,\;\; p<0,\\
-1, & t_{\Gamma}<t_0,\;\; p\ge 0,\\
3-\dfrac{2}{w_1+1}, & t_{\Gamma}>t_0,
\end{cases}
\label{eq:freqscaling}
\end{equation}
where $t_\Gamma$ denotes the characteristic time at which GW emission becomes important for the loop evolution and $p=-3+2/(1+w_1)+4/(3(1+w_2))$. The parameter $p$ characterizes which emission times dominate the integral for $C_m$. For $p<0$, the dominant contribution comes from loops emitting near $t_\Gamma$, whereas for $p\ge 0$ it comes from the latest possible emission time in the corresponding cosmological era. In the standard cosmological history, this reproduces a distinct power law spectrum
\begin{equation}
\Omega_{\rm GW}^{(1)}(f)\propto
\begin{cases}
f^{0}, & \text{formed: RD, emitted: RD},\\
f^{-1/2}, & \text{formed: RD, emitted: MD},\\
f^{-1}, & \text{formed: MD, emitted: MD}.
\end{cases}
\end{equation}
Moreover, the higher harmonics obey
\begin{equation}
\Omega_{\rm GW}^{(m)}(f)=m^{-q}\,\Omega_{\rm GW}^{(1)}(f/m),
\end{equation}
so the total spectrum after summing over all $m$ can be modified relative to the $m=1$ estimate, especially around spectral breaks. These standard-cosmology results serve as the reference spectrum against which non-standard expansion histories can be compared.

\subsection{GW spectrum modified by matter and kination eras}\label{subsec:modified cosmology}

The important point for the present work is that cosmic strings are long lasting sources of GW. Because of this, the observed frequency today carries information about both the epoch when the relevant loops were formed and the epoch when they emit GW. Any non-standard expansion can leave an imprint on the string induced SGWB, e.g., Refs.~\cite{Gouttenoire:2019kij, Ghoshal:2023sfa, Eto:2024hwn,Gouttenoire:2021wzu}.

For sufficiently long MD and KD eras, the string spectrum is modified into a triangular shape.
If we neglect the sum over higher harmonics, one finds the following asymptotic behavior
\begin{equation}
\Omega_{\rm GW}(f)\propto
\begin{cases}
f^0, & f \lesssim f_{\rm KR}, \\
f^{+1}, & f_{\rm KR}\lesssim f \lesssim f_{\rm MK}, \\
f^{-1}, & f_{\rm MK}\lesssim f \lesssim f_{\rm RM}, \\
f^0, & f \gtrsim f_{\rm RM},
\end{cases}
\label{eq:piecewise_string_singlemode}
\end{equation}
where $f_{\rm KR}$, $f_{\rm MK}$, and $f_{\rm RM}$ denote the characteristic frequencies associated with the transitions at $T_{\rm KR}$, $T_{\rm MK}$, and $T_{\rm RM}$, respectively. The origin of these slopes follows directly from Eq.~(\ref{eq:freqscaling}). The segment $f_{\rm KR}\lesssim f \lesssim f_{\rm MK}$ probes loops associated with the kination era. Since kination is a stiff phase ($w=1$), the background energy density redshifts faster than radiation, so the GW contribution from earlier times is weighted more strongly toward higher observed frequencies, which leads to $\Omega_{\rm GW}\propto f^{+1}$. The segment $f_{\rm MK}\lesssim f \lesssim f_{\rm RM}$ instead probes loops associated with the preceding MD. In that epoch, the expansion history shifts the emission frequency relation in the opposite direction, leading to a negative spectral tilt towards higher frequencies, which leads to $\Omega_{\rm GW}\propto f^{-1}$. As a result, the standard broad plateau is converted into a peaked structure around the matter-to-kination transition.

For realistic local strings, however, the observed spectrum is the sum over all harmonics. For cusp-dominated loops, this leaves the kination slope essentially unchanged, while the high frequency tail associated with the preceding matter era is substantially flattened. In that case, Eq.~\eqref{eq:piecewise_string_singlemode} is replaced more accurately by
\begin{equation}
\Omega_{\rm GW}(f)\propto
\begin{cases}
f^0, & f \lesssim f_{\rm KR}, \\
f^{+1}, & f_{\rm KR}\lesssim f \lesssim f_{\rm MK}, \\
f^{-1/3}, & f_{\rm MK}\lesssim f \lesssim f_{\rm RM}, \\
f^0, & f \gtrsim f_{\rm RM},
\end{cases}
\label{eq:piecewise_string_total}
\end{equation}
where the $f^{-1/3}$ is due to cusp-dominated emission after summing over higher modes~\cite{Blasi:2020wpy}.

\begin{figure*}
    \centering
    \includegraphics[width=\textwidth]{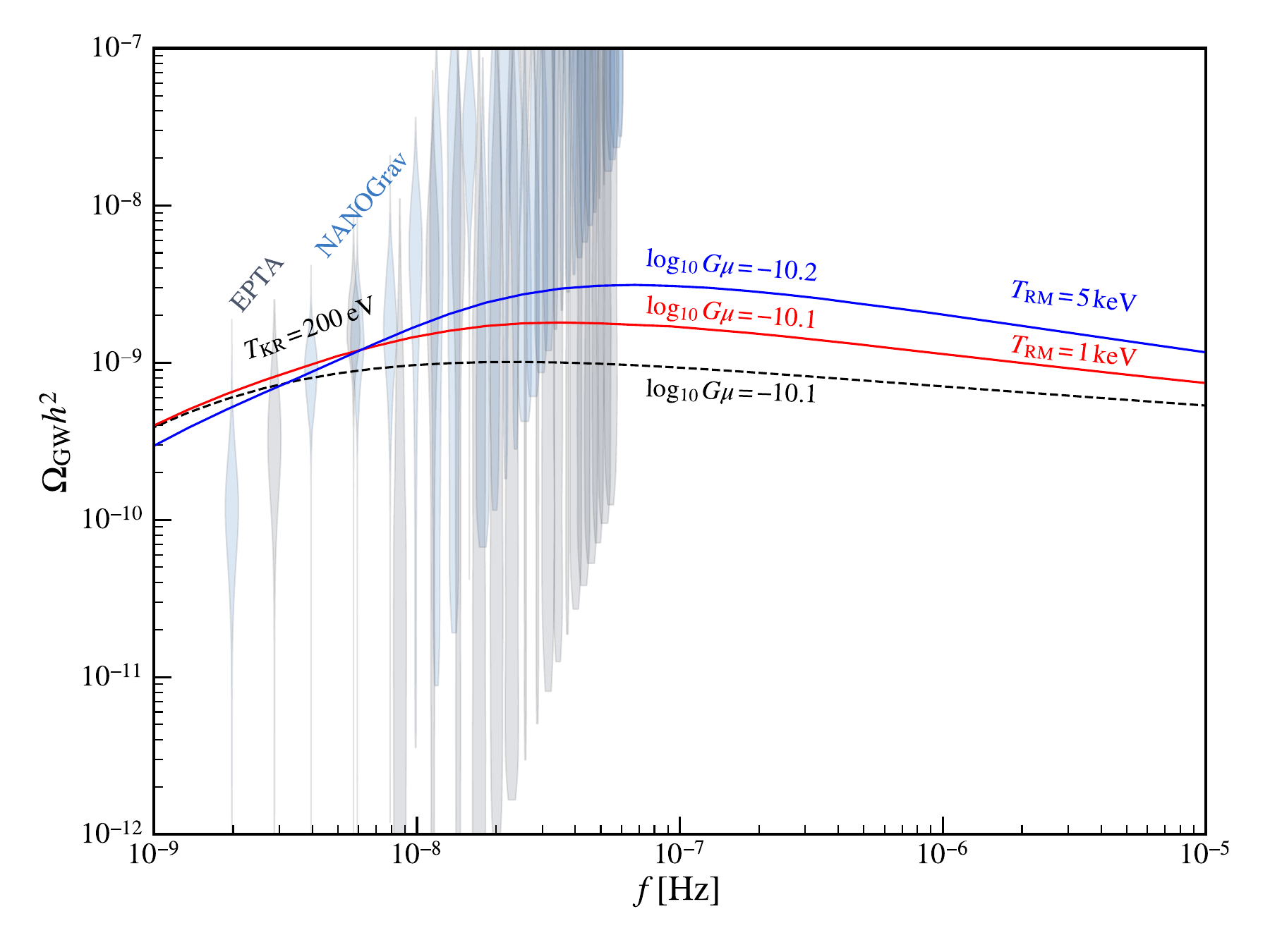}
    \caption{Illustrative SGWB spectra generated from cosmic strings in the kination cosmology compared to the counterpart in $\Lambda$CDM. The black dashed line denotes the spectrum in $\Lambda$CDM with the string tension fixed at $\log_{10}G\mu=-10.1$. The red and blue solid lines show the spectra in the presence of a matter-kination epoch for $\log_{10}G\mu=-10.1$ with $T_{\rm RM}=1 \;\rm keV$ and $\log_{10}G\mu=-10.2$ with $T_{\rm RM}=5 \;\rm keV$. We fix $T_{\rm KR}=200\;\rm eV$, $\alpha=0.1$, $F=0.1$, $\Gamma=50$, and $q=4/3$, and sum over $10^4$ modes. The violins represent the HD free-spectrum result from NANOGrav 15-year and EPTA DR2 New+ datasets.}
    \label{fig:GWSpectrum_PTA_kination_demo}
\end{figure*}

Hence, in the axion rotation cosmology discussed in Sec.~\ref{sec:kination}, the local string SGWB acquires a characteristic triangular bump or peaked broken power law.
Its location is determined primarily by $T_{\rm MK}$ through $f_{\rm MK}$, while its width is controlled by the duration of the non-standard era, namely by the hierarchy between $T_{\rm RM}$ and $T_{\rm KR}$. This feature is especially important for the present work. Because the cosmic string signal is intrinsically broad, a non-standard era does not create an isolated narrow peak from scratch, but rather reshapes the broad radiation-era plateau into a triangular spectrum. Consequently, a measurement of the break frequencies and slopes can in principle probe not only the microscopic string parameters $(G\mu,\alpha,\Gamma)$ but also the axion rotation parameters encoded in $T_{\rm RM}$, $T_{\rm MK}$, and $T_{\rm KR}$. The effect of kination cosmology on the cosmic-string-induced SGWB is illustrated in Fig.~\ref{fig:GWSpectrum_PTA_kination_demo}. The solid contour lines show $\Omega_{\rm GW}h^2$ in the kination cosmology for different combinations of \{${\log_{10}G\mu, T_{\rm RM}}$\}. We fix $T_{\rm KR}=200\;\rm eV$, $\alpha=0.1$, $F=0.1$, $\Gamma=50$, and $q=4/3$, and sum over $10^4$ modes. In contrast, the dashed black line represents a typical SGWB from cosmic strings in $\Lambda$CDM with $\log_{10}G\mu$ fixed at $-10.1$. The comparison demonstrates that an epoch of kination can significantly enhance the spectral peak of the cosmic-string-induced SGWB.

\section{Bayesian analysis framework}\label{sec:bayesian}
In this section, we discuss the main components of statistical analysis within the Bayesian framework, establishing a systematic approach for evaluating model evidence and, consequently, the Bayes factor for model selection. For a more comprehensive review of Bayesian analysis and model selection, we refer the reader to the literature, e.g., Refs.~\cite{10.1093/oso/9780198568315.001.0001, James:2006zz}.

For a model $M$ and data $D$, the posterior is obtained from the Bayes' theorem:
\begin{equation}\label{eq:posterior}
  \mathcal{P}(\boldsymbol{\theta}|D,M) = \frac{{\pi(\boldsymbol{\theta}|M)}\mathcal{L}(D|\boldsymbol{\theta},M)}{\mathcal{Z}(D|M)},  
\end{equation}
where 
\begin{itemize}
    \item $\pi(\boldsymbol{\theta}|M)$ is the prior of the parameter $\boldsymbol{\theta}$ assuming model $M$, encoding all prior information before observing the data $D$.
    \item $\mathcal{L}(D|\boldsymbol{\theta},M)$ denotes the likelihood of data $D$ given $M$ and $\boldsymbol{\theta}$.
    \item $\mathcal{Z}(D|M)$ is the Bayesian evidence, which constitutes a crucial quantity for model comparisons. This can be obtained by integrating the likelihood over the parameter space weighted by the prior 
    \begin{equation}\label{eq:evidence}
        \mathcal{Z}(D|M)=\int{\mathcal{L}(D|\boldsymbol{\theta},M) \pi(\boldsymbol{\theta}|M)d\boldsymbol{\theta}}.
    \end{equation}
\end{itemize} 

The Bayes factor $B_{ij}$ is used to compare two competing models, $M_i$, $M_j$ with the dataset $D$, within the Bayesian framework and is defined as the ratio of the respective Bayesian evidence calculated from Eq.~(\ref{eq:evidence}),
\begin{equation}\label{eq:Bayes_factor}
    B^{D}_{ij}=\frac{\mathcal{Z}(D|M_i)}{\mathcal{Z}(D|M_j)}.
\end{equation}
If $B^{D}_{ij} > 1$, there is evidence in favor of model $M_i$ over model $M_j$; if $B^{D}_{ij} < 1$, then $M_j$ is preferred over $M_i$. In this work, we quote the log Bayes factor, $\ln B^{D}_{ij}=\ln {\mathcal Z}^{D}_i - \ln \mathcal {Z}^{D}_j$, and use Jeffreys’ scale~\cite{Jeffreys:1939xee, Robert:2008vdd} to interpret the relative performance of any two competing models $M_i$ and $M_{j}$. The linear and log Jeffreys' scales are summarized in Table~\ref{tab:Jeffreys' scale}. We explore the parameter space for each model and the Bayesian evidence using \texttt{UltraNest}~\cite{Buchner:2021cql}. The marginalized posteriors presented in this work are derived from the nested samples using kernel density estimations implemented in the \texttt{GetDist} package~\cite{Lewis:2019xzd}.

\begin{table*}[htbp]
\centering \renewcommand{\arraystretch}{1.3} \begin{tabular}{|p{2.3cm}|p{2.5cm}|p{3cm}|} \hline $B_{ij}$ & $\ln B_{ij}$ & Interpretation  \\ 
\hline
$<1$ & $< 0$ & $M_j$ is favored\\
\hline
(1, $10^{1/2}$) & (0,1.2) & Negligibly small \\
\hline
($10^{1/2}$, 10) & (1.2, 2.3) & Substantial \\
\hline
(10, $10^{3/2}$) & (2.3, 3.5) & Strong \\
\hline 
($10^{3/2}$, $10^{2}$) & (3.5, 4.6) & Very strong \\
\hline
$>10^2$& $> 4.6 $ & Decisive\\
\hline
\end{tabular}
\caption{Linear and logarithmic Jeffreys' scales and the corresponding interpretation}
\label{tab:Jeffreys' scale}
\end{table*}

Bayesian evidence naturally incorporates the principle of Occam’s razor, penalizing models with unnecessary complexity. Consequently, one can extract information about "goodness-of-fit" and the model complexity from Bayes evidence. This can be achieved by decomposing log evidence into two terms:
\begin{equation}\label{eq:evidence decomposition}
    \ln\mathcal{Z}^{D}_M = \langle\ln\mathcal{L^{D}} \rangle_{\mathcal{P},M} - \mathcal{D}^{D}_{\mathrm{KL},M} .
\end{equation}
The first term is the posterior-weighted log likelihood, defined as
\begin{equation}\label{eq:posterior-weighted likelihood}
    \langle\ln\mathcal{L}^{D} \rangle_{\mathcal{P},M}
    = \int \ln \mathcal{L}(D|\boldsymbol{\theta},M)\mathcal{P}(\boldsymbol{\theta}|D,M) \rm{d} \theta,
\end{equation}
which measures how well the model fits the data, averaged over the posterior distribution, with larger values meaning the posterior concentrates on parameter values that explain the data better. The second term is the Kullback–Leibler (KL) divergence~\cite{Kullback:1951zyt,Handley:2019pqx}, defined as
\begin{equation}\label{eq:KL divergence}
    \mathcal{D}^{D}_{\mathrm{KL},M}
    = \int \mathcal{P}(\boldsymbol{\theta}|D,M)
    \ln\frac{\mathcal{P}(\boldsymbol{\theta}|D,M)}{\pi(\boldsymbol{\theta}|M)} \rm{d}\theta,
\end{equation}
which quantifies the overall compression of the posterior relative to the prior, weighted by the posterior distribution. This term represents the effective Occam penalty: models for which the data restrict the posterior to only a small fraction of the prior space will have a larger KL divergence. The decomposition of the log evidence using Eqs.~\eqref{eq:evidence decomposition}, \eqref{eq:posterior-weighted likelihood} and \eqref{eq:KL divergence} enables contributions from model fit and the effective model complexity to be assessed separately across different models 
\begin{equation}\label{eq:Bayes_factor_decomposition}
\ln {B}^{D}_{ij}
=
\Delta\langle \ln\mathcal{L}^{D}\rangle_{P,ij}
-
\Delta \mathcal{D}^{D}_{\mathrm{KL},ij},
\end{equation}
where $\Delta X_{ij} = X_i - X_j$. We use the \texttt{Anesthetic} package~\cite{Handley:2019mfs} to compute these quantities.

\section{Analyses}
\label{sec:Analyses}
In this section, we describe the setup adopted for the Bayesian analysis. Our main goal is to determine how well the cosmic string induced SGWB can fit PTA data in the presence of an intermediate non-standard cosmological era. To address this question, we consider a modified expansion history between BBN and recombination, following the sequence
\begin{equation*}
{\rm RD}\rightarrow{\rm MD}\rightarrow{\rm KD}\rightarrow{\rm RD}.
\end{equation*}
As discussed in Sec.~\ref{sec:CS_GW}, such a non-standard history deforms the broad cosmic string spectrum into a broken power-law shape with a characteristic bump. If the transition scales are between BBN and recombination, this feature can be shifted into the nanohertz frequency range probed by PTA experiments, where it may provide a better fit to the observed signal than the prediction from standard cosmology. This motivates the study of not only the PTA fit, but also whether the corresponding non-standard era remains consistent with BBN and the CMB.

In realistic UV complete models~\cite{Co:2019wyp}, the transitions between these epochs are smooth. In the present analysis, however, we adopt a simplified description in which the background evolution is treated piecewise with instantaneous transitions in the scaling of the scalar field's energy density with respect to the scale factor. This approximation captures the leading impact of the non-standard era on the cosmic string gravitational wave spectrum while simplifying the numerical calculation for the Bayesian analysis. The expansion history is therefore specified by the characteristic transition temperatures $T_{\rm RM}$ and $T_{\rm KR}$ introduced in the previous sections with $T_{\rm MK}$ automatically determined by Eq.~(\ref{MKrelation}). These parameters determine the duration and timing of the MD and KD, and hence control the corresponding deformation of the cosmic string spectrum relevant for PTA observations. Consequently, we compute the time evolution of the scale factor $a(t)$ from the first Friedmann equation 
    \begin{align}\label{eq:Friedmann}
        H^2=H_0^2\left[\Delta_r(a)\Omega_{r}\left(\frac{a}{a_0}\right)^{-4}+\Omega_{m}\left(\frac{a}{a_0}\right)^{-3}+\Omega_{\Lambda}+
    \Omega_{k,\theta}\left(\frac{a}{a_{\mathrm{KR}}}\right)^{-6}+\Omega_{m,\theta}\left(\frac{a}{a_{\mathrm{MK}}}\right)^{-3}\right],
    \end{align}
where $\Omega_{m,\theta}=\Omega_r(\frac{a_{\mathrm{KR}}}{a_0})^{-4}(\frac{a_{\mathrm{MK}}}{a_{\mathrm{KR}}})^{-6}\Theta(a_{\mathrm{MK}}-a)$ and $\Omega_{k,\theta}=\Omega_r(\frac{a_{\mathrm{KR}}}{a_0})^{-4}\Theta(a-a_{\mathrm{MK}})$ are, respectively, the density parameters of the rotating condensate during early matter-dominated and kination-dominated eras relative to the current critical energy density $\rho_{c,0} =\frac{3H_0^2}{8\pi G}$. $\Omega_r=8.88\times10^{-5}$, $\Omega_m=0.30$, $\Omega_\Lambda=1-\Omega_r-\Omega_m$ denote the present-day density parameters for radiation, matter and dark energy, respectively, and the Hubble constant is $H_0=68.6 \; \mathrm{km\; s^{-1} \;Mpc^{-1}}$. These values are obtained from the Bayesian inference which will be described in Sec.~\ref{subsubsec:CMB}. The multiplicative factor 
\begin{equation}
    \Delta_r(a) = \frac{g_{*}(a)}{g_{*}(a_0)}\left(\frac{g_{*s}(a_0)}{g_{*s}(a)}\right)^{4/3}
\end{equation}
is inserted to account for the departure from the relation $a\propto T^{-1}$ owing to entropy conservation and depends upon the relativistic degrees of freedom for which we use the tabular data computed in Ref.~\cite{Saikawa:2020swg}. The case of $\Omega_{k,\theta} = \Omega_{m, \theta}=0$ in Eq.~(\ref{eq:Friedmann}) corresponds to $\Lambda \mathrm{CDM}$. In the following sections, we examine the constraints from BBN and the CMB on the non-standard cosmology parameters $T_{\rm RM}$ and $T_{\rm KR}$ in Sec.~\ref{subsec:constraints}, followed by PTA analysis in Sec.~\ref{subsec: PTAs and analyses Setup}. 

\subsection{Constraints}
\label{subsec:constraints}
Our inference proceeds in three stages. The non-standard expansion history is parametrized by the radiation-to-matter and kination-to-radiation transition temperatures, $T_{\rm RM}$ and $T_{\rm KR}$, with $T_{\rm MK}$ determined by Eq.~\eqref{MKrelation}. First, we use BBN light-element abundances to constrain $T_{\rm RM}$ and construct its marginalized posterior distribution. We then fix $T_{\rm RM}$ to the $2\sigma$ BBN upper bound, together with the corresponding helium abundance, and use CMB and baryon acoustic oscillations (BAO) observations to constrain $T_{\rm KR}$. The prior range for $T_{\rm KR}$ is chosen to ensure that the Universe returns to radiation domination before recombination while maintaining the required ordering $T_{\rm KR}\leq T_{\rm RM}$. As shown below, the inferred $T_{\rm KR}$ distribution is largely insensitive to variations of $T_{\rm RM}$ within its BBN-allowed range, motivating the separate BBN and CMB+BAO fits. Finally, kernel-density estimates of the marginalized BBN and CMB+BAO posteriors are used as informative priors for $T_{\rm RM}$ and $T_{\rm KR}$, respectively, in the PTA analysis, where both transition temperatures are varied together with the cosmic-string and astrophysical parameters. Thus, BBN and CMB+BAO first determine the allowed non-standard expansion histories, while the NANOGrav and EPTA data subsequently test their imprint on the gravitational-wave spectrum.

\subsubsection{BBN}
\label{subsubsec:BBN}
An early matter domination will enhance the expansion rate. For a transient matter domination at a temperature of the order $\mathcal{O}(\rm{keV})$, observations of primordial helium and deuterium abundances provide a sensitive probe to constrain $T_{\rm{RM}}$ as faster expansion shortens the duration of the nuclear-burning phase, thereby altering the primordial light-element abundances.

\begin{figure*}[t!]
    \centering
    \includegraphics[width=\columnwidth]{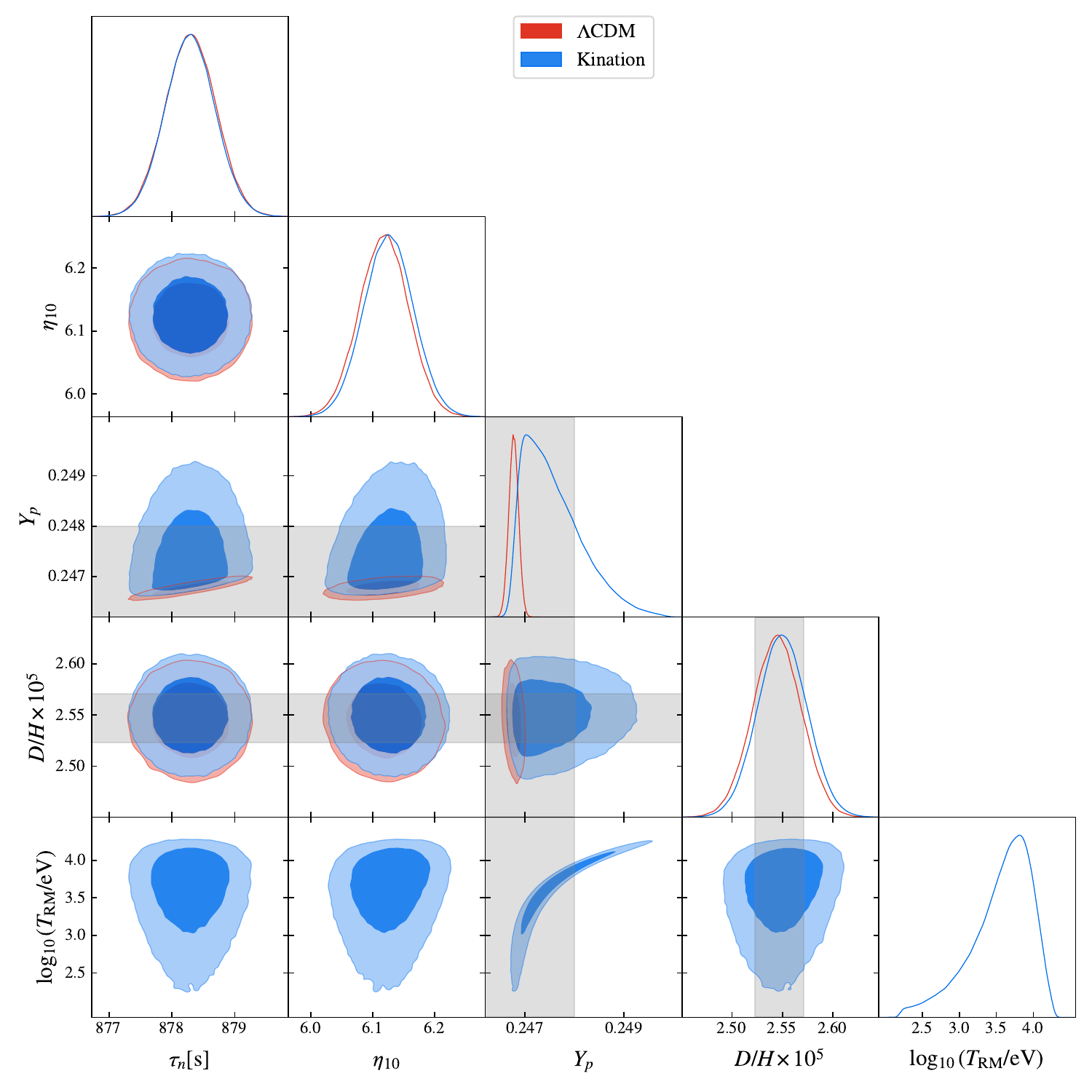}
    \caption{Marginalized posterior distributions (68\% and 95\% C.L.) for $\log_{10}T_{\rm{RM}}$ with $T_{\rm{KR}}$ fixed to 150 eV, neutron lifetime $\tau_n$, baryon-to-photon ratio $\eta_{10}$ and the derived helium and deuterium abundances $Y_p$, $D/H\times10^5 $ in $\Lambda$CDM (red) and kination (blue). The grey bands correspond to $Y_p$ and $D/H$ measurements~\cite{Aver:2021rwi, Fernandez:2019hds, Aver:2020fon, Hsyu:2020uqb}.}
    \label{fig:BBN_posterior}
\end{figure*}

To study the effect of a brief departure from standard cosmology on light-element abundances, we modify the publicly available BBN solver \texttt{PRyMordial}~\cite{Burns:2023sgx} to include the axion energy density in the thermal background. We adopt a Gaussian prior $\mathcal{N}(\bar{\eta}_{10}, \sigma_{\eta_{10}}^2)$ on the baryon-to-photon ratio, parametrized as $\eta_{10}=10^{10}\eta_B$, where $\bar{\eta}_{10}$ and $\sigma_{\eta_{10}}$ are converted from Planck-inferred $\Omega_bh^2$~\cite{Planck:2018vyg}. We adopt the Particle Data Group~\cite{ParticleDataGroup:2020ssz} average neutron lifetime $\tau_n=878.3\pm0.4$ s, derived from eight measurements made with ultracold neutrons~\cite{Musedinovic:2024gms, Ezhov:2014tna, Pattie:2017vsj, Serebrov:2017bzo, Arzumanov:2015tea, Steyerl:2012zz, Pichlmaier:2010zz, Serebrov:2004zf} and impose a Gaussian prior $\tau_n\sim \mathcal{N} (878.3, 0.4^2)$. Through \texttt{smallnet=True} flag in the PRyMordial configuration file, we marginalize over twelve key nuclear reaction rates from NACRE II compilation~\cite{Xu:2013fha}, $\vec{\Gamma}=\left(\Gamma_1, \Gamma_2,...,\Gamma_{12}\right)$, that are necessary to accurately predict the primordial helium and deuterium abundances. We model each nuclear rate as log-normal distributed:
\begin{equation}
    \log\Gamma_i = \log\langle\Gamma_{i}\rangle +{\beta_i} \log\left(\Delta\Gamma_i\right),
\end{equation}  
where $\beta_i$ is the nuisance parameter governing the variation of $i$-th nuclear reaction rate sampled from a normal distribution, $\langle\Gamma_i\rangle$ denotes the median rate, and $\Delta \Gamma_i$ is the corresponding uncertainty factor. We define the log likelihood function 
\begin{equation}\label{eq:BBN_likelihood}
    \log\mathcal{L} = -\frac{(Y_{p}-Y^{\rm obs}_{p})^2}{2\sigma_{Y_p^{\rm obs}}^2}-\frac{((D/H)-(D/H)^{\rm obs})^2}{2\sigma_{(D/H)^{\rm obs}}^2},
\end{equation}
where $Y_p$ and $D/H$ are the predicted primordial helium mass fraction and deuterium abundance derived from \texttt{PRyMordial}, $Y_p^{\rm obs}=0.245$ and $(D/H)^{\rm obs}=2.547\times10^{-5}$ are the central values of light abundances from different measurements~\cite{Aver:2021rwi, Fernandez:2019hds, Aver:2020fon, Hsyu:2020uqb} and $\sigma_{Y_p^{\rm obs}}=0.003$ and $\sigma_{(D/H)^{\rm obs}}=0.025\times10^{-5}$ are their respective uncertainties.

We use \texttt{emcee}~\cite{Foreman-Mackey:2012any}, an affine-invariant ensemble sampler~\cite{Goodman:2010dyf}, to derive the posterior distributions for the 15-dimensional parameter space ($T_{\rm{RM}}, \eta_{10}, \tau_n,\vec{\beta}$) with $T_{\rm{KR}}$ fixed to $150\;{\rm eV}$. We run $N_{\rm walker}=128$ for 50000 iterations each. The first 35\% of each chain is discarded as burn-in. We assess convergence by first finding the maximum integrated autocorrelation time $\tau_{f}^{\rm max}$ across all parameters. We then obtain an estimate of the effective sample size ${\rm ESS} = N_{\rm walker}\times N_{\rm s}/{\tau_f^{\rm max}}$, where $N_{\rm s}$ is the number of samples of each walker after burn-in. This yields an ${\rm ESS} \approx 1.8\times 10^{4}$. The one- and two-dimensional marginalized posterior distributions are shown in Fig.~\ref{fig:BBN_posterior}. We find the 68\% and 95\% C.L. upper bounds on $T_{\rm RM}$ as 
\begin{align}\label{eq:TRM_1sigma}
    T_{\rm RM} & \leq 6.45\;\rm keV, \\
\label{eq:TRM_2sigma}
    T_{\rm RM} & \leq 12.6\;\rm keV .
\end{align}

In the joint CMB+BAO analysis presented in Sec.~\ref{subsubsec:CMB}, when we impose the temperature ordering $T_{\rm KR} \leq T_{\rm RM}$, we adopt the 2$\sigma$ value of $T_{\rm RM}$ as the upper bound on $T_{\rm KR}$. The banana-shape  in the $T_{\rm{RM}}$-$Y_p$ posterior reveals a strong correlation between the two parameters. This results in a stringent constraint on $T_{\rm{RM}}$ from BBN. To obtain a representative $Y_p$ at the 2$\sigma$(1$\sigma$) upper bound of $T_{\rm RM}$, we select samples within a finite interval centered on the 2$\sigma$($1\sigma$) upper bound of $T_{\rm RM}$ and take the corresponding median of $Y_p$. This procedure yields $Y_p\sim0.2487(0.2478)$, which we adopt as an input to the Boltzmann solver \texttt{CLASS}. The $T_{\rm RM}$ samples provide an informative prior for the GW background analysis in Sec.~\ref{subsec: PTAs and analyses Setup}. 

\subsubsection{CMB}
\label{subsubsec:CMB}

The presence of a kination epoch can affect the CMB anisotropies and CMB lensing power spectrum. As discussed in Ref.~\cite{Co:2021lkc}, the modified expansion rate can alter the evolution of cosmological perturbations and reduce the comoving sound horizon $r_s$. This modifies the acoustic scale and can also alter the matter power spectrum, potentially leaving observable signatures in structure-growth and distance-scale probes such as CMB lensing and BAO. To quantify these effects and constrain the non-standard cosmology parameters, we perform a joint likelihood analysis of the CMB and BAO data using \texttt{CLASS}~\cite{Blas:2011rf}, modified to incorporate the non-standard cosmological history described earlier, together with the \texttt{Cobaya} sampler~\cite{Torrado:2020dgo,2019ascl.soft10019T}. We take six $\Lambda \textrm{CDM}$ parameters and one non-standard cosmology parameter $T_{\rm KR}$,
\begin{equation}
\{\omega_b,\omega_c,H_0,\ln(10^{10}A_s),n_s,\tau_{\rm reio},T_{\rm KR}\},
\end{equation}
as free parameters while fixing $T_{\rm RM}$ to a value near the upper bound allowed by BBN at the $2\sigma$ level in Eq.~(\ref{eq:TRM_2sigma}), by which we require $T_{\rm KR}$ to be bounded from above. The parameter priors used in the CMB analysis are summarized in Table~\ref{tab:CMB priors}. Since $T_{\rm RM}$ is strongly correlated with the primordial helium abundance $Y_p$ (Fig.~\ref{fig:BBN_posterior}), we determine the median value of $Y_p$ for each fixed $T_{\rm RM}$. The resulting pair is $(T_{\rm RM},Y_p)=(12.6\;{\rm keV},0.2487)$, which is adopted as an input for the background cosmology in \texttt{CLASS}. We verify that the posterior distribution of $T_{\rm KR}$ is largely insensitive to variations in $T_{\rm RM}$. We therefore perform separate fits to the BBN and CMB data.

\begin{table*}[htbp]
    \centering
    \setlength{\tabcolsep}{18pt}
    \begin{tabular}{lc}
    \hline
    Parameter & Prior \\
    \noalign{\hrule height 1.0pt}
    $ \omega_b$ & $\mathcal{U}(0.020, 0.025)$ \\
    $ \omega_c$ & $\mathcal{U}(0.095, 0.145)$ \\
    $ H_0$ & $\mathcal{U}(60, 75)$ \\
    $ \ln(10^{10}A_s)$ & $\mathcal{U}(2.5, 3.9)$ \\
    $ n_s$ & $\mathcal{U}(0.86, 1.06)$ \\
    $ \tau_{\rm reio}$ & $\mathcal{U}(0.01, 0.15)$ \\
    $ \log_{10}(T_{\rm KR}/\rm eV) $ & $\mathcal{U}(0, 4.10)$ \\
    \hline
    \end{tabular}
    \caption{Prior used in the CMB analysis. Here $\mathcal{U}$ denotes a uniform distribution.}
    \label{tab:CMB priors}
\end{table*}

\begin{figure*}[t!]
    \centering
    \includegraphics[width=\columnwidth]{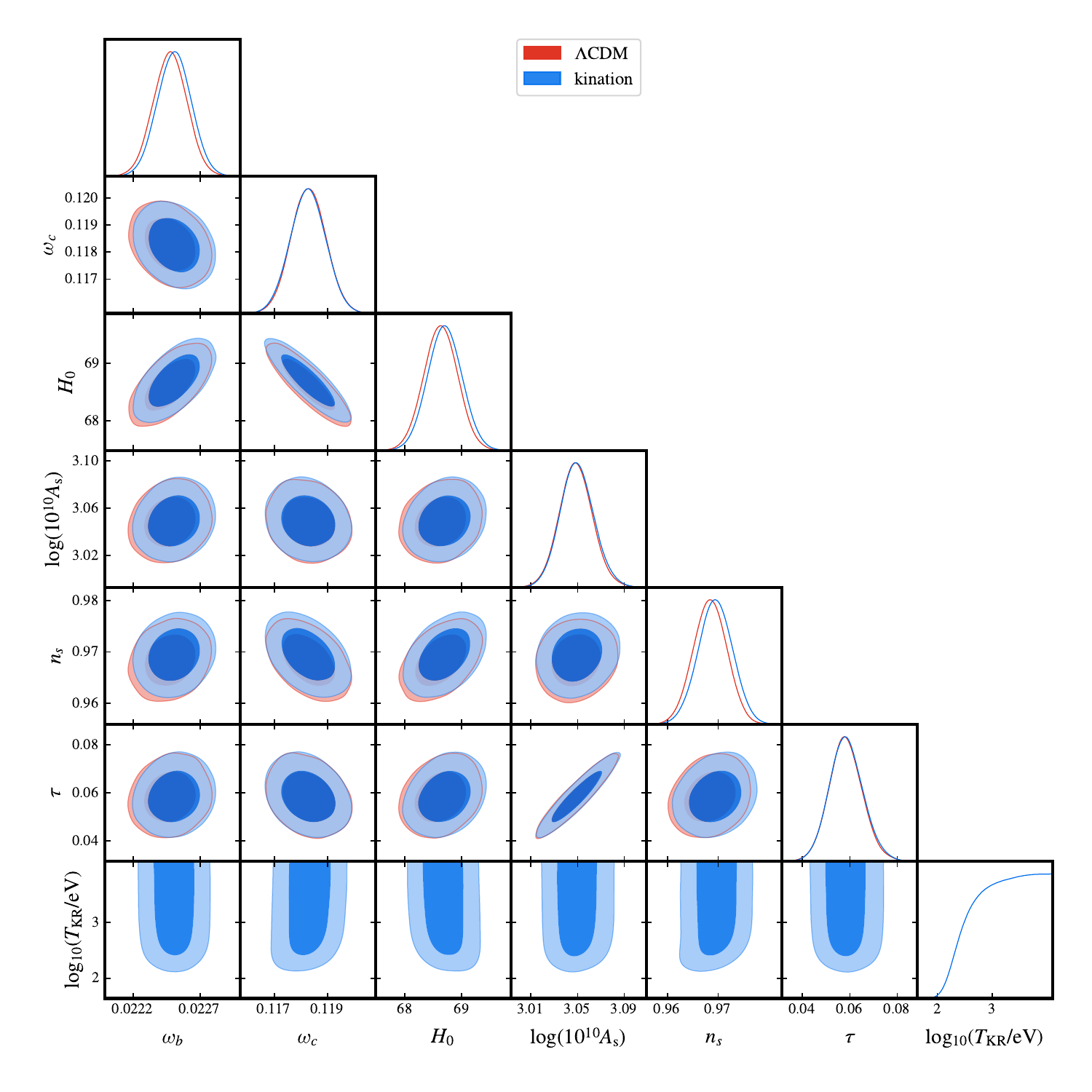}
    \caption{Marginalized posterior distributions showing 68\% and 95\% regions for cosmological parameters and $\log_{10}(T_{\rm{KR}}/\rm{eV})$ for $\Lambda$CDM (red) and kination (blue) CMB (including {Planck} 2018 low-$\ell$ TT, EE, high-$\ell$ TTTEEE and lensing likelihoods) and DESI DR2 BAO. For kination, $T_{\rm{RM}}$ is fixed to 12.6 keV and $Y_p$ is fixed to 0.2487.}
    \label{fig:cmb_posterior}
\end{figure*}

For the CMB likelihoods, we use the Planck 2018 temperature and polarization power spectra, including the low-$\ell$ TT, EE, high-$\ell$ TTTEEE likelihoods~\cite{Planck:2018vyg}, and CMB lensing~\cite{Planck:2018lbu}. For the BAO likelihood, we include BAO measurements from DESI DR2~\cite{DESI:2025zgx}. The posterior is sampled using MCMC implemented in \texttt{Cobaya}, and convergence is declared once the Gelman--Rubin criterion~\cite{Gelman:1992zz} satisfies $|R-1|<0.005$. The posterior distribution for the cosmological parameters is shown in Fig.~\ref{fig:cmb_posterior}.

\subsection{PTAs and analyses setup}
\label{subsec: PTAs and analyses Setup}
A passing gravitational wave perturbs the spacetime metric along the line of sight between millisecond pulsars and Earth, which induces a time-dependent perturbation in pulse time-of-arrival measurements. This effect can be parametrized by a one-sided timing-residual power spectral density (PSD) $S(f_i)={\Phi(f_i)}T_{\rm{obs}}$ at each frequency bin $f_i=i/T_{\rm{obs}}$, where $T_{\rm{obs}}$ is the observation baseline and $i$ is the harmonic number. The fractional GW energy density is related to the PSD via the relation 
\begin{equation}
    \Omega_{\mathrm{GW}}(f)h^2 = \frac{8\pi^4f^5}{H_{100}^2}S(f).
\end{equation}

In this work, we perform separate Bayesian analyses of the NANOGrav 15-year (NG15) and the EPTA Data Release 2 New+ (EPTA DR2 New+) datasets, using the probability density distributions inferred from the Hellings–Downs (HD) correlated free-spectrum analyses from the respective datasets~\cite{the_nanograv_collaboration_2023_10344086, EPTA:2023akd, antoniadis_2023_8091568}. The EPTA DR2 New+ dataset combines a 10.3-year subset of the EPTA observations obtained with improved observing systems with the data from the Indian Pulsar Timing Array (InPTA) dataset~\cite{Tarafdar:2022toa}. For NG15, we use the publicly available kernel density estimation (KDE)-based likelihood representation released through \texttt{PTArcade}~\cite{Mitridate:2023oar}. For EPTA DR2 New+, we use the publicly available \texttt{Ceffyl} package~\cite{Lamb:2023jls} to construct a KDE-based representation of the likelihood. The total log-likelihood is given by 
$$
\ln \mathcal{L}^{D} = \sum_{i}^n\ln\mathcal{L}_i^{{D}}(\vec{\theta}),
$$
where $D$ labels the dataset and $i$ represents the bin index. We restrict our analysis to the first 14 and the 10 frequency components in NG15 and EPTA DR2 New+, respectively. These likelihoods are implemented in \texttt{UltraNest}. 

\noindent
\textbf{Baseline model: Supermassive black hole binaries---}
As our baseline astrophysical interpretation of the nanohertz gravitational wave background, we consider a population of SMBHBs. SMBHBs are one of the most popular astrophysical sources of an SGWB in the nanohertz frequency regime. This is mainly because SMBHBs are expected to be abundant in the Universe, and because the GWs emitted during their inspiral can easily fall within the frequency band probed by pulsar timing arrays.

The GW spectrum from SMBHBs can be parametrized as~\cite{Ellis:2020ena}
\begin{equation}\label{eq:GW_spectgrum_SMBHB}
    \Omega_{\rm GW} h^2 =
    \frac{2\pi^2 f_\textrm{yr}^{2} A_{\rm SMBHB}^2}{3H_{100}^2}
    \left(\frac{f}{f_\textrm{yr}^{}}\right)^{5-\gamma_{\rm SMBHB}}
    ,
\end{equation}
where $H_{100}=100\;{\rm km\,s^{-1}\,Mpc^{-1}}$, $f_{\rm{yr}}=1/\rm{yr}$, $A_{\rm SMBHB}$ is the GW amplitude from the SMBHB population, and $\gamma_{\rm SMBHB}$ is the power law spectral index. The amplitude is primarily determined by the masses of the SMBHBs and by the abundance of SMBHBs contributing to the population. On the other hand, the spectral index encodes information about how SMBHBs evolve before and during the GW driven inspiral phase.

Currently, both the amplitude and the spectral index remain weakly constrained by observations and numerical simulations. The current constraint is obtained from astrophysical simulations using the \texttt{holodeck} package~\cite{NANOGrav:2023hfp}, which implements semi-analytic modeling of SMBHB population. For our baseline hypothesis $\mathcal{H}_0$, we therefore adopt a simulation-informed prior on the parameters $(\log_{10}A_{\rm SMBHB}, \gamma_{\rm SMBHB})$. Specifically, they follow a two-dimensional Gaussian distribution whose mean and covariance matrix are obtained from fitting to numerical simulations~\cite{NANOGrav:2023hvm},
\begin{equation}\label{eq:prior_SMBHB}
  \boldsymbol{\mu}_{\rm SMBHB}=
\begin{pmatrix}-15.6\\4.7\end{pmatrix},
\qquad
\boldsymbol{\Sigma}_{\rm SMBHB}=
\begin{pmatrix}0.28&-0.0026\\-0.0026&0.12\end{pmatrix}.  
\end{equation}

\begin{table*}[t]
\centering \renewcommand{\arraystretch}{1.3} \begin{tabular}{|p{1.5cm}|p{3cm}|p{4cm}|p{4cm}|} 
\hline Model & Descriptions & Parameters & Prior  \\ 
\hline
$\mathcal{H}_0$ & SMBHBs  & $\left(\log_{10}A_{\rm{SMBHB}} ,\gamma_{\rm{SMBHB}}\right)$ & $\mathcal{N}(\boldsymbol{\mu}_{\rm{SMBHB}}, \boldsymbol{\Sigma}_{\rm{SMBHB}})$ \\
\hline
$\mathcal{H}_1$ & CS & $\log_{10}G\mu$ &  $\mathcal{U}(-14, -6)$\\
\hline
$\mathcal{H}_2$  & CS+SMBHBs & $\log_{10}G\mu$ & $\mathcal{U}(-14, -6)$ \\
                &    & $\left(\log_{10}A_{\rm{SMBHB}},\gamma_{\rm{SMBHB}}\right)$ & $\mathcal{N}(\boldsymbol{\mu}_{\rm{SMBHB}}, \boldsymbol{\Sigma}_{\rm{SMBHB}})$ \\
\hline
$\mathcal{H}_3$ & kin-CS & $\log_{10}G\mu$ & $\mathcal{U}(-14, -6)$ \\
                &        &  $\log_{10}(T_{\mathrm{RM}}/{\rm eV})$ & BBN posterior samples \\
                &        &  $\log_{10}(T_{\mathrm{KR}}/{\rm eV})$ & CMB posterior samples \\
\hline
$\mathcal{H}_4$ & kin-CS+SMBHBs & $\log_{10}G\mu$ & $\mathcal{U}(-14, -6)$ \\
                &        &  $\log_{10}(T_{\mathrm{RM}}/{\rm eV})$ &  BBN posterior samples\\
                &        &  $\log_{10}(T_{\mathrm{KR}}/{\rm eV})$ & CMB posterior samples \\
                &        & $\left(\log_{10}A_{\rm{SMBHB}},\gamma_{\rm{SMBHB}}\right)$ & $\mathcal{N}(\boldsymbol{\mu}_{\rm{SMBHB}}, \boldsymbol{\Sigma}_{\rm{SMBHB}})$ \\
\hline
\end{tabular}
\caption{A summary of the five hypotheses analyzed in this work and their model parameters, together with the prior ranges adopted in the SGWB analysis. Here $\mathcal{U}$ and $\mathcal{N}$ denote uniform and normal distributions, respectively. The quantities $\mu_{\rm{SMBHB}}$ and $\Sigma_{\rm{SMBHB}}$ denote the two-dimensional Gaussian prior mean vector and covariance matrix determined from simulation in Ref.~\cite{NANOGrav:2023hvm} and given in Eq.~\eqref{eq:prior_SMBHB}. The informative priors for $\log_{10}(T_{\rm{RM}}/\rm{eV})$ and $\log_{10}(T_{\rm{KR}}/\rm{eV})$ are respectively constructed from the posterior samples obtained in the BBN and CMB analyses described in Sec.~\ref{subsubsec:BBN} and Sec.~\ref{subsubsec:CMB}, respectively.}
\label{tab:model description}
\end{table*}

\noindent
\textbf{Alternative hypotheses: cosmic strings---}
We consider four alternative hypotheses: cosmic-string models within either $\Lambda$CDM or the modified cosmology (kin) described in Sec.~\ref{subsec:modified cosmology}, with SMBHB contributions (CS+SMBHBs, kin-CS+SMBHBs) or without them (CS, kin-CS). The descriptions and parameters of each model are summarized in Table~\ref{tab:model description}. For all alternative models, the cosmic-string GW spectrum in Eq.~\eqref{eq:GW_spectrum_CS} is truncated to the first $10^4$ modes\footnote{Higher harmonics contribute to higher frequency GW as can be seen from Eq.~\eqref{eq:freqharmonic}. We truncate the harmonic sum at $10^4$ beyond which we find no appreciable change in the predicted GW spectrum in the nanohertz regime.} and we assume that the GW power spectrum is dominated by cusps ($q=4/3$). For the kination parameters, we use the posterior samples of $\log_{10}(T_{\rm{RM}}/\rm{eV})$ and $\log_{10}(T_{\rm{KR}}/\rm{eV})$ from the independent BBN and CMB analyses directly as informative priors. We implement prior transformations that map unit cube variables to $\log_{10}{T_{\rm RM}}$ and $\log_{10}{T_{\rm KR}}$ in \texttt{UltraNest} according to the empirical posterior distribution encoded in the corresponding BBN and CMB chains. We adopt a stopping criterion of \texttt{frac\_remain}=0.01 and require a minimum of \texttt{nlive}=400 live points and an effective sample size of at least 1000 to declare convergence. 

\section{Results}\label{sec:results}
We now present the results of our Bayesian analyses of the NG15 and EPTA DR2 New+ datasets. To isolate the impact of the modified expansion history associated with kination, we first compare the cosmic-string models in $\Lambda$CDM and kination cosmologies, taking the SMBHB-only model as the common baseline $\mathcal{H}_0$. We then perform the same comparison for models that include an additional SMBHB contribution. Finally, we directly compare the $\Lambda$CDM and kination scenarios within the cosmic-string framework, both with and without an additional SMBHB contribution.

\begin{table*}[t]
\centering
\renewcommand{\arraystretch}{1.3}
\begin{tabular}{|p{1.5cm}|p{2.5cm}|p{2.5cm}|p{2.5cm}|p{2.5cm}|p{2.5cm}|}
\hline
Model & $\log_{10}G\mu$ & $\mathrm{log}_{10}(T_{\rm{RM}}/\rm{eV})$ & $\log_{10}(T_{\rm{KR}}/\rm{eV})$ & $\log_{10}A_{\rm{SMBHB}}$ & $\gamma_{\rm{SMBHB}}$ \\
\hline
$\mathcal{H}_1 $ & $-10.10^{+0.21}_{-0.24}$ & - & - & - & - \\
$\mathcal{H}_2 $ & $-11.26^{+1.07}_{-1.84}$ & - & - & $-14.64^{+0.18}_{-0.85}$& $4.27^{+0.42}_{-0.33}$ \\
$\mathcal{H}_3 $ & $-10.15^{+0.17}_{-0.17}$ & $ 3.82^{+0.23}_{-0.32}$  & $2.49^{+0.54}_{-0.25}$ & - & - \\
$\mathcal{H}_4 $ & $-10.45^{+0.36}_{-2.28}$ & $3.79^{+0.25}_{-0.34}$  & $2.79^{+0.61}_{-0.45}$ & $-14.90^{+0.40}_{-1.00}$ & $4.38^{+0.44}_{-0.39}$ \\
\hline
\end{tabular}
\caption{Summary of 68\% credible intervals for the model parameters of different hypotheses fit to the NG15 dataset.}
\label{tab:model_summary_ng15}

\vspace{0.5cm}

\centering
\renewcommand{\arraystretch}{1.3}
\begin{tabular}{|p{1.5cm}|p{2.5cm}|p{2.5cm}|p{2.5cm}|p{2.5cm}|p{2.5cm}|}
\hline
Model & $\log_{10}G\mu$ & $\mathrm{log}_{10}(T_{\rm{RM}}/\rm{eV})$ & $\log_{10}(T_{\rm{KR}}/\rm{eV})$ & $\log_{10}A_{\rm{SMBHB}}$ & $\gamma_{\rm{SMBHB}}$ \\
\hline
$\mathcal{H}_1 $ & $ -9.99^{+0.25}_{-0.24} $ & - & - & - & - \\
$\mathcal{H}_2 $ & $-10.16^{+0.30}_{-1.62}$ & - & - & $-15.31^{+0.53}_{-0.72}$ & $4.66^{+0.37}_{-0.35}$ \\
$\mathcal{H}_3 $ & $ -10.03^{+0.23}_{-0.23}$ & $  3.77^{+0.25}_{-0.35}$  & $2.99^{+0.51}_{-0.48}$ & - & - \\
$\mathcal{H}_4 $ & $-10.16^{+0.29}_{-1.23}$ & $3.76^{+0.25}_{-0.32}$  & $3.03^{+0.50}_{-0.49}$ & $-15.44^{+0.62}_{-0.59} $ & $4.66^{+0.34}_{-0.35}$ \\
\hline
\end{tabular}
\caption{Summary of 68\% credible intervals for the model parameters of different hypotheses fit to the EPTA DR2 New+ dataset.}
\label{tab:model_summary_epta}
\end{table*}

\begin{table*}[t]
    \centering
    \setlength{\tabcolsep}{18pt}
    \begin{tabular}{ccc}
    \hline
    \noalign{\vskip 3.5pt}
    Model Comparison
    & $\ln{B}_{i0}^{\rm NG15}\;(\pm\sigma)$ & $\ln{B}_{i0}^{\rm EPTA}\;(\pm\sigma)$ \\
    \noalign{\hrule height 1.0pt}
    $\mathcal{H}_1$ vs. $\mathcal{H}_0$
        & $-1.74 \pm 0.07$  & $0.14\pm0.06$\\
        $\mathcal{H}_2$ vs.\ $\mathcal{H}_0$
        & $-0.50 \pm 0.08$  & $0.47\pm0.09$\\
        $\mathcal{H}_3$ vs.\ $\mathcal{H}_0$
        & $-0.45 \pm 0.07$ & $0.28\pm0.07$\\
        $\mathcal{H}_4$ vs.\ $\mathcal{H}_0$
        & {$0.25\pm 0.11$}  & {$0.63\pm0.08$}\\
    \hline
    \end{tabular}
    \caption{Natural log Bayes factors comparing different SGWB source models given the NANOGrav 15-year and EPTA DR2 New+ datasets under the baseline SMBHB hypothesis $\mathcal{H}_0$. A log Bayes factor with $\ln B^{D}_{i0}>0$ indicates a preference over the reference model $\mathcal{H}_0$. }
    \label{tab:bayes_factors}

\vspace{0.5cm}

    \centering
    \setlength{\tabcolsep}{15pt}
    \begin{tabular}{ccc}
    \hline
    \noalign{\vskip 3pt}
    \multirow[b]{2}{*}{Model Comparisons} & \multicolumn{2}{c}{$\ln B^D_{ij}$} \\
    
    \cline{2-3}
     & $D$=NG15 & $D$=EPTA \\
     \noalign{\hrule height 1.0pt}
     $\mathcal{H}_3$ vs $\mathcal{H}_1$ &$1.29\pm0.06 $& $0.14\pm0.06$\\
     $\mathcal{H}_4$ vs $\mathcal{H}_2$ & $0.75\pm 0.11 $& $0.16\pm 0.10$\\
    \hline
\end{tabular}
\caption{Natural log Bayes factors comparing models with kination and their counterpart in $\Lambda$CDM. $\ln B^{D}_{ij}>0$ indicates a preference for model \textit{i} over model \textit{j} with the dataset $D$.}
\label{tab:bayes-lcdm-vs-kination}
\end{table*}

The inferred parameter constraints for all models are summarized in Tables~\ref{tab:model_summary_ng15} and~\ref{tab:model_summary_epta} for the NG15 and EPTA DR2 New+ datasets, respectively. The corresponding Bayes factors relative to the baseline hypothesis $\mathcal{H}_0$ are reported in Table~\ref{tab:bayes_factors}. The direct comparisons between kination and $\Lambda$CDM within the cosmic-string framework with and without SMBHBs are summarized in Table~\ref{tab:bayes-lcdm-vs-kination}. All quoted parameter constraints correspond to 68\% credible intervals.

\subsection{Bayesian inference and model comparison}
\subsubsection{Results for NG15 dataset}

\begin{figure*}[t!]
    \centering
    \includegraphics[width=0.7\columnwidth]{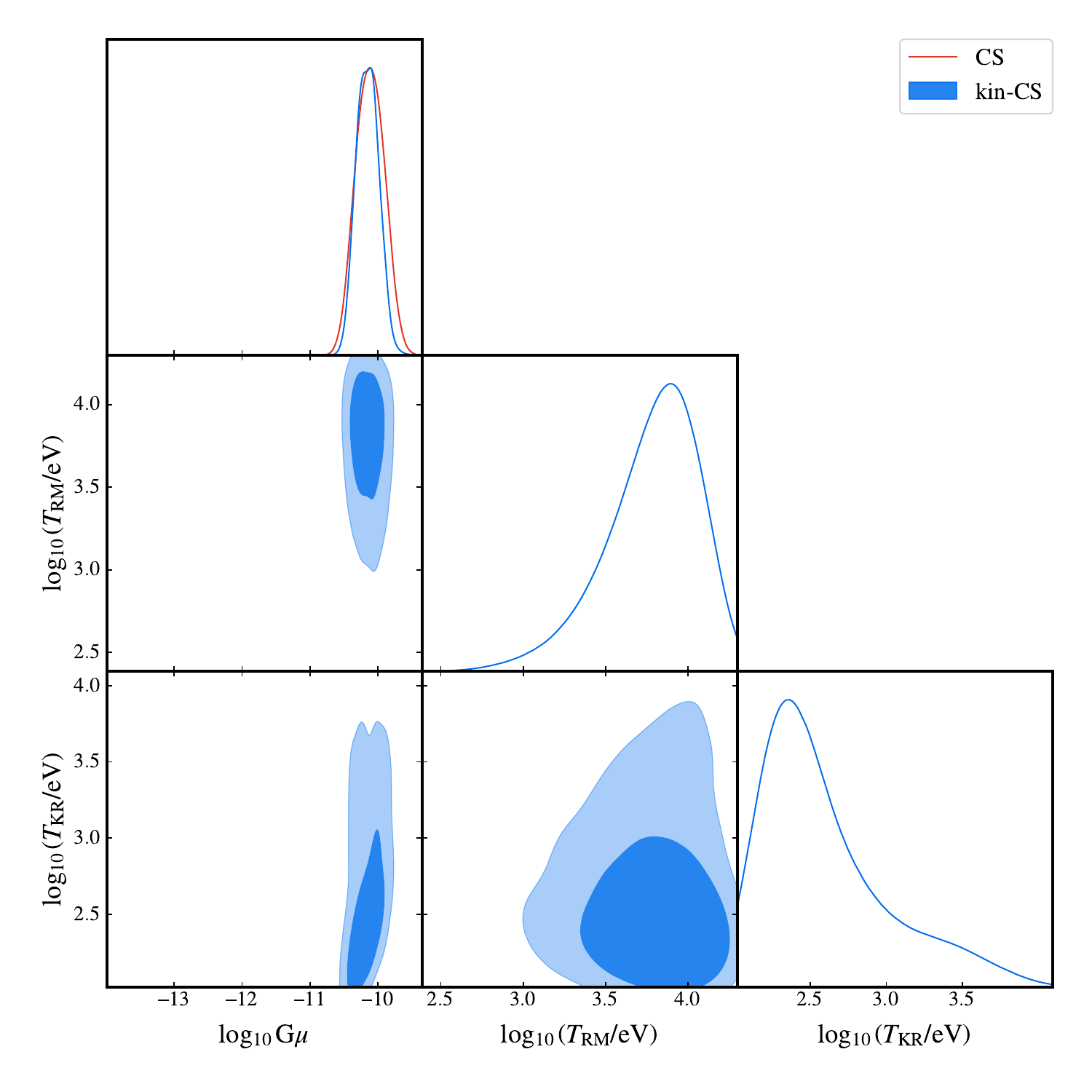}
    \caption{Marginalized posterior distributions (68\% and 95\% C.L.) for cosmic-string-only hypotheses $\mathcal{H}_1$ and $\mathcal{H}_3$ in $\Lambda$CDM (red) and kination (blue). Constraints are derived from fits to NG15 discussed in Sec.~\ref{subsec: PTAs and analyses Setup}}
    \label{fig:cs_posterior_ng15}
\end{figure*}

\textbf{CS: $\Lambda$CDM  vs kination---}
The posterior distributions of the cosmic-string model in $\Lambda$CDM $\mathcal{H}_1$ and in the presence of a matter-kination epoch $\mathcal{H}_3$ obtained from the NG15 analysis are shown in Fig.~\ref{fig:cs_posterior_ng15}. We find $\log_{10}G\mu=-10.10^{+0.21}_{-0.24}$ for $\Lambda$CDM while $\log_{10}G\mu=-10.15^{+0.17}_{-0.17}$ for the kination scenario. For the latter, the two additional transition temperatures $\log_{10}(T_{\rm{RM}}/\rm{eV})$ and $\log_{10}(T_{\rm{KR}}/\rm{eV})$ are constrained to $ \log_{10}(T_{\rm RM}/\rm{eV})=3.82^{+0.23}_{-0.32}$  and $\log_{10}(T_{\rm{KR}}/\rm{eV})=2.49^{+0.54}_{-0.25}$, corresponding to the credible intervals $T_{\rm RM} \in [3.1,11.2 ]\; \rm{keV}$ and $T_{\rm KR}\in [174,1077]\; \rm{eV}$. The Bayes factors relative to the baseline SMBHB model $\mathcal{H}_0$ are
\begin{equation}\label{eq:bayes factor - CS_ng15}
    \ln {B}^{\rm NG15}_{10}= -1.74 \pm 0.07,
\end{equation}
\begin{equation} \label{eq:bayes factor - kin-CS_ng15}
    \ln {B}^{\rm NG15}_{30} = -0.45 \pm 0.07.
\end{equation}

Although both hypotheses $\mathcal{H}_1$ and $\mathcal{H}_3$ have lower Bayesian evidence than $\mathcal{H}_0$, introducing a transient matter-kination epoch substantially reduces the evidence deficit relative to the baseline $\mathcal{H}_0$ while leaving the marginalized posterior distributions of $\log_{10}G\mu$ in the two scenarios virtually invariant. 

\begin{figure*}[t!]
    \centering
    \includegraphics[width=\columnwidth]{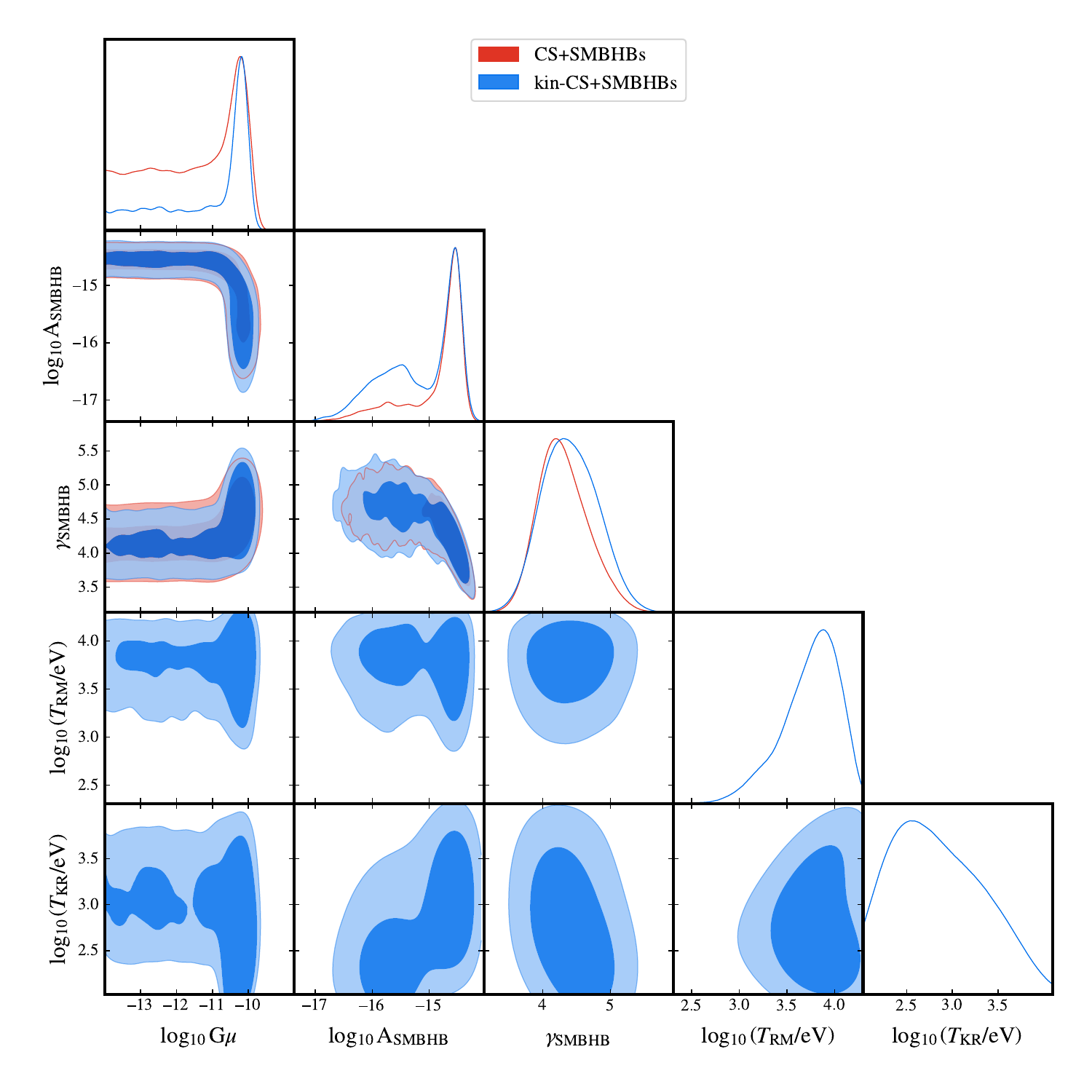}
    \caption{Marginalized posterior distributions (68\% and 95\% C.L.) for cosmic-string hypotheses with SMBHB contributions, $\mathcal{H}_2$ and $\mathcal{H}_4$, in $\Lambda$CDM (red) and kination cosmology (blue). Constraints are derived from fits to NG15 discussed in Sec.~\ref{subsec: PTAs and analyses Setup}}
    \label{fig:cs+smbhb_posterior_ng15}
\end{figure*}
\noindent
\textbf{CS+SMBHBs: $\Lambda$CDM  vs kination---}
We next present the posterior distributions and Bayes factors when contributions from SMBHBs are superimposed on those of cosmic strings in the two cosmological scenarios. The posterior distributions of hypotheses $\mathcal{H}_2$ and $\mathcal{H}_4$ are shown in Fig.~\ref{fig:cs+smbhb_posterior_ng15}. The inferred cosmic-string tension in $\mathcal{H}_2$ is $\log_{10}G\mu=-11.26^{+1.07}_{-1.84}$ while that in $\mathcal{H}_4$ is $\log_{10}G\mu=-10.45^{+0.36}_{-2.28}$. For SMBHB model parameters, the amplitude and slope in $\mathcal{H}_2$ are constrained at $\log_{10}A_{\rm SMBHB} =-14.64^{+0.18}_{-0.85}$ and $\gamma_{\rm SMBHB} = {4.27}^{+0.42}_{-0.33}$ whereas for $\mathcal{H}_4$ the constraints are $\log_{10}A_{\rm SMBHB} = -14.90^{+0.40}_{-1.00}$ and $\gamma_{\rm SMBHB}=4.38^{+0.44}_{-0.39}$. For the two non-standard cosmology parameters in $\mathcal{H}_4$, we find $\log_{10}(T_{\rm{RM}}/\rm{eV})=3.79^{+0.25}_{-0.34}$ and $\log_{10}(T_{\rm{KR}}/\rm{eV})=2.79^{+0.61}_{-0.45}$, corresponding to the intervals $T_{\rm RM}\in[2.82, 10.90]\;\rm{keV}$ and $T_{\rm KR}\in[222, 2537]\;\rm{eV}$. Including an SMBHB contribution reduces the difference in log evidence between the cosmic-string models and the SMBHB-only baseline $\mathcal{H}_0$:
\begin{align}\label{eq:bayes factor - CS_smbhb_ng15}
    \ln {B}^{\rm NG15}_{20} &= -0.50 \pm 0.08, \\
\label{eq:bayes factor - kin-CS_smbhb_ng15}
    \ln {B}^{\rm NG15}_{40} &= 0.25 \pm 0.10. 
\end{align}

\subsubsection{Results for EPTA DR2 New+ dataset}

\begin{figure*}
    \centering
    \includegraphics[width=0.7\columnwidth]{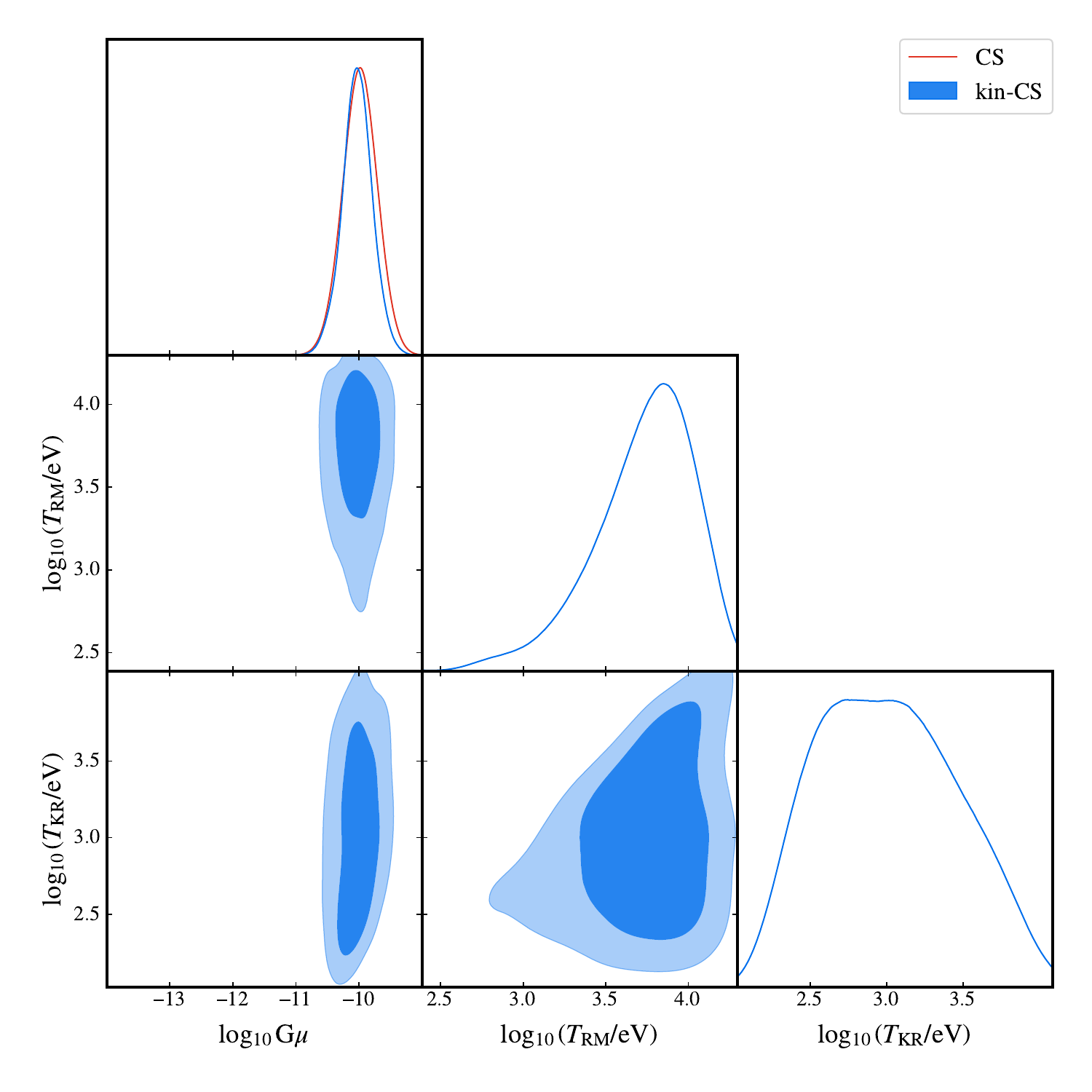}
    \caption{Marginalized posterior distributions (68\% and 95\% C.L.) for cosmic-string-only hypotheses, $\mathcal{H}_1$ and $\mathcal{H}_3$, in $\Lambda$CDM (red) and kination (blue), obtained from fits to EPTA DR2 New+ discussed in Sec.~\ref{subsec: PTAs and analyses Setup}}
    \label{fig:cs_posterior_epta}
\end{figure*}

\textbf{CS: $\Lambda$CDM  vs kination---}
The posterior distributions of hypotheses $\mathcal{H}_1$ and $\mathcal{H}_3$ fitted to EPTA DR2 New+ are shown in Fig.~\ref{fig:cs_posterior_epta}. For model parameters, we obtain $\log_{10}G\mu=-9.99^{+0.25}_{-0.24} $ for $\mathcal{H}_1$ and $\log_{10}G\mu=-10.03^{+0.23}_{-0.23}$ for $\mathcal{H}_3$. The constraints on transition temperatures are $\log_{10}(T_{\rm{RM}}/\rm{eV})=3.77^{+0.25}_{-0.35}$ and $\log_{10}(T_{\rm{KR}}/\rm{eV})=2.99^{+0.51}_{-0.48}$, yielding $T_{\rm RM}\in[2.6, 10.4]\;\rm{keV}$ and $T_{\rm KR}\in[329, 3180]\;\rm{eV}$ while the Bayes factors of the two hypotheses compared to $\mathcal{H}_0$ are calculated as 
\begin{align}\label{eq:bayes factor - CS_epta}
   \ln {B}^{\rm EPTA}_{10} &=  0.14\pm 0.06,  \\
   \label{eq:bayes factor - kin-CS_epta}
    \ln {B}^{\rm EPTA}_{30} &= 0.28 \pm 0.07. 
\end{align}
The best estimates of the log Bayes factors of both hypotheses are positive but remain close to zero, indicating a mild preference over $\mathcal{H}_0$. In addition, introducing kination slightly improves the posterior-averaged log-likelihood, but the improvement is partially offset by an increase in the KL divergence, resulting in a net positive change in the log Bayes factor.

\begin{figure*}[t!]
    \centering
    \includegraphics[width=\columnwidth]{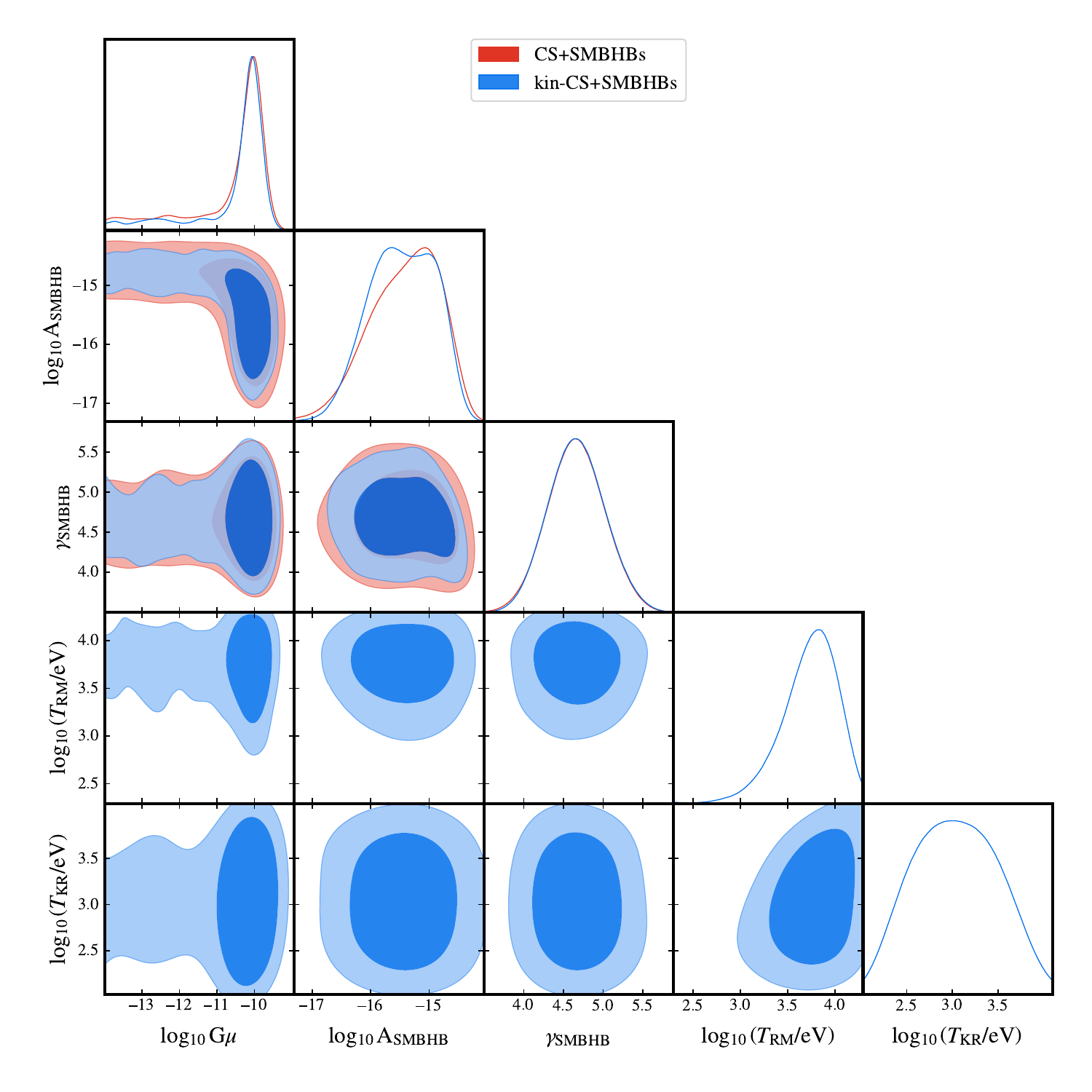}
    \caption{Marginalized posterior distributions (68\% and 95\% C.L.) for cosmic-string hypotheses with SMBHB contributions, $\mathcal{H}_2$ and $\mathcal{H}_4$, in $\Lambda$CDM (red) and kination cosmology (blue), obtained from fits to EPTA DR2 New+ discussed in Sec.~\ref{subsec: PTAs and analyses Setup}}
    \label{fig:cs+smbhb_posterior_epta}
\end{figure*}
\noindent
\textbf{CS+SMBHBs: $\Lambda$CDM  vs kination---}
The posterior distributions of hypotheses $\mathcal{H}_2$ and $\mathcal{H}_4$ fitted to EPTA DR2 New+ are shown in Fig.~\ref{fig:cs+smbhb_posterior_epta}.
Their Bayes factors relative to $\mathcal{H}_0$ are estimated as
\begin{align}\label{eq:bayes factor - CS_smbhb_epta}
    \ln {B}^{\rm EPTA}_{20} & = 0.47 \pm 0.09, \\
\label{eq:bayes factor - kin-CS_smbhb_epta}
    \ln {B}^{\rm EPTA}_{40} & = 0.63 \pm 0.08.
\end{align}

The inclusion of an SMBHB contribution alongside cosmic strings slightly increases the relative log Bayes factor in both the $\Lambda$CDM and kination scenarios. For the derived constraints in $\mathcal{H}_2$ and $\mathcal{H}_4$, we find that the string tension peaks at similar values in both scenarios $\log_{10}G\mu=-10.16^{+0.30}_{-1.62}$ and $\log_{10}G\mu=-10.16^{+0.29}_{-1.23}$, respectively. For the astrophysical parameters, we find $\log_{10}A_{\rm SMBHB}=-15.31^{+0.53}_{-0.72}$ and $\gamma_{\rm SMBHB}=4.66^{+0.37}_{-0.35}$ for $\mathcal{H}_2$ and $\log_{10}A_{\rm SMBHB}=-15.44^{+0.62}_{-0.59}$ and $\gamma_{\rm SMBHB}=4.66^{+0.34}_{-0.35}$ for $\mathcal{H}_4$. The two temperature parameters in $\mathcal{H}_4$ are constrained at $\log_{10}(T_{\rm{RM}}/\rm{eV})=3.76^{+0.25}_{-0.32}$ and $\log_{10}(T_{\rm{KR}}/\rm{eV})=3.03^{+0.50}_{-0.49}$. In linear units, these correspond to $T_{\rm RM}\in [2.8, 10.3]\; \rm keV$ and $T_{\rm KR} \in [349, 3394]\;\rm eV$.

\subsubsection{Model comparison}
The log Bayes factors in Eqs.~\eqref{eq:bayes factor - CS_ng15}-\eqref{eq:bayes factor - kin-CS_smbhb_epta} can be decomposed, based on Eq.~\eqref{eq:Bayes_factor_decomposition}, in the form of $\ln {B}^{D}_{ij} =\Delta\langle \ln\mathcal{L}^{D}\rangle_{P,ij} - \Delta \mathcal{D}^{D}_{\mathrm{KL},ij}$. The decomposition shows whether a change in Bayesian model support originates from an improved posterior-weighted log likelihood or from the corresponding KL divergence. The log evidence decompositions of all models relative to $\mathcal{H}_0$ are summarized in Table~\ref{tab:Bayes_factor_decomposition}. 
\begin{table*}[htbp]
    \centering
    \begin{tabular}{lcccc}
        \hline
        Model comparison 
        & $\Delta\langle\ln L^{\rm NG15}\rangle_{P,i0}$ 
        & $\Delta \mathcal{D}^{\rm NG15}_{\mathrm{KL},i0}$ 
        & $\Delta\langle\ln L^{\rm EPTA}\rangle_{P,i0}$ 
        & $\Delta \mathcal{D}^{\rm EPTA}_{\mathrm{KL},i0}$ 
        \\
        \hline
        $\mathcal{H}_1-\mathcal{H}_0$ &$-4.41\pm0.03$ &$-2.66\pm0.06$ & $-0.29\pm0.03$&$-0.43\pm0.06$ \\
        $\mathcal{H}_2-\mathcal{H}_0$ & $-1.39\pm0.09$& $-0.89\pm0.07$&$-0.12\pm0.04$ & $-0.59\pm0.09$\\
        $\mathcal{H}_3-\mathcal{H}_0$ & $-2.05\pm0.04$ & $-1.60\pm0.07$& $0.03\pm0.03$& $-0.25\pm0.06$ \\
        $\mathcal{H}_4-\mathcal{H}_0$ &$-1.12\pm0.09$ & $-1.37\pm0.06$& $0.07\pm0.03$&$-0.57\pm0.07$ \\
        \hline
    \end{tabular}
    \caption{Decomposition of log Bayes factor comparing the alternative hypotheses against baseline model across the NG15 and EPTA datasets. The first and second columns under each dataset represent the posterior-weighted likelihood and KL divergence term.}
    \label{tab:Bayes_factor_decomposition}
\end{table*}

For NG15 dataset analysis, none of the alternative hypotheses yields a better posterior-weighted log likelihood compared to SMBHBs. Despite being close to zero, $\mathcal{H}_4$ is the only alternative hypothesis favored over the SMBHB-only baseline $\mathcal{H}_0$, owing to its more favorable KL-divergence contribution. For the EPTA DR2 New+ dataset, $\mathcal{H}_3$ and $\mathcal{H}_4$ yield marginally higher log likelihood terms than $\mathcal{H}_0$; however, the improvements in the Bayes factors are primarily driven by the more favorable KL-divergence contributions.

We next assess the effect of kination by directly comparing the kination hypothesis with its $\Lambda$CDM counterpart within the cosmic-string framework. For the cosmic-string-only model, this corresponds to comparing $\mathcal{H}_3$ with $\mathcal{H}_1$, while for models including SMBHBs, we compare $\mathcal{H}_4$ with $\mathcal{H}_2$. The corresponding log Bayes factors are summarized in Table~\ref{tab:bayes-lcdm-vs-kination}. The direct comparison between $\mathcal{H}_3$ and $\mathcal{H}_1$ in NG15 dataset indicates substantial support for kination over $\Lambda$CDM, yielding $\ln B^{\rm NG15}_{31}=1.29\pm0.06$, or equivalently $B^{\rm NG15}_{31} = 3.63^{+0.22}_{-0.21}$. This comparison is particularly relevant if the SMBHB contribution to the SGWB is small, as hinted at by astrophysical population simulations, in which case the PTA signal may be predominantly sourced by cosmic strings. The role of kination cosmology is apparent; the presence of a brief matter-kination epoch leads to a spectral enhancement over its counterpart in $\Lambda$CDM through the modified expansion history, improving the fit sufficiently to overcome the associated increase in the KL penalty and consequently increasing the log Bayes factor. This improvement occurs because the GW spectrum depends on the expansion history of the universe when the cosmic-string loops emit GWs. As discussed in Sec.~\ref{subsec:modified cosmology}, a brief period of matter-kination epoch modifies the expansion history and hence the relation between loop formation and emission times and the observed GW frequency, altering the spectral shape relative to $\Lambda$CDM. As shown by Eq.~\eqref{eq:piecewise_string_total}, these effects produce a blue-tilted spectrum, $\Omega_{\rm GW}\propto f$, for loops associated with the kination era, enhancing the higher-frequency part of the spectrum relative to $\Lambda$CDM, despite the nearly unchanged constraints on cosmic-string tension. This enhancement allows the model to better accommodate the blue-tilted $\Omega_{\rm GW}h^2$ spectrum inferred from the NG15 data. This preference, however, is reduced once contributions from SMBHBs are included. In contrast, EPTA shows little ability to distinguish between the two cosmologies, with kination yielding a slightly positive log Bayes factor.

\subsection{Gravitational-wave spectra}
The SGWB spectra for all hypotheses and PTA dataset considered in this work are presented in this section. For each hypothesis, parameter samples are redrawn from the posterior and mapped onto the corresponding GW energy-density spectrum, $\Omega_{\rm GW}h^2$. The median spectrum and 68\% credible interval are determined from the resulting ensemble of spectral realizations at each frequency. For comparison, we show the SMBHB contributions to the SGWB denoted by a black dashed line in each plot throughout this section, adopting the fiducial values $\log_{10}A_{\rm SMBHB}=-15.6$ and $\gamma_{\rm SMBHB}=4.7$, obtained from astrophysical simulations~\cite{NANOGrav:2023hvm}.

\begin{figure*}
    \centering
    \includegraphics[width=\textwidth]{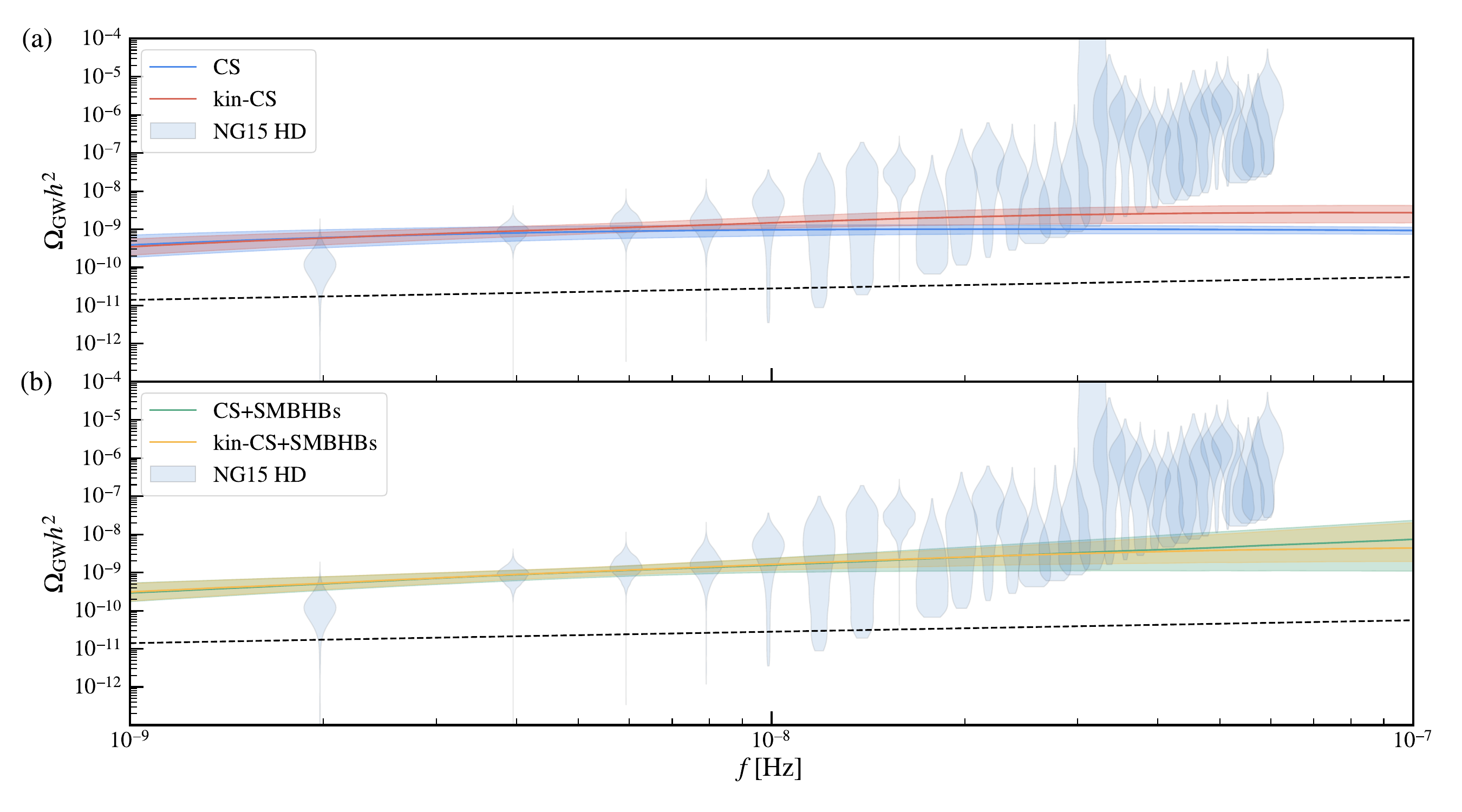}
    \caption{The median gravitational wave spectra induced by the four alternative hypotheses for the NG15 dataset. The blue violins represent the HD-correlated free spectra. (a) Cosmic-string spectra in $\mathrm{\Lambda}$CDM (CS, blue) and kination (kin-CS, red) cosmologies. (b) The corresponding spectra superimposed with an additional SMBHB contribution (CS+SMBHBs, green; kin-CS+SMBHBs, orange). The shaded bands indicate the 68\% credible intervals.  The black dashed lines represent the GW signal from SMBHBs with $\log_{10}A_{\rm SMBHB}=-15.6$ and $\gamma_{\rm SMBHB}=4.7$, which are the mean values from astrophysical simulations~\cite{NANOGrav:2023hvm}.}
    \label{fig:median_GWSpectrum_ng15}
\end{figure*}

The SGWB spectra of the alternative models fitted to the NG15 dataset are shown in Fig.~\ref{fig:median_GWSpectrum_ng15}. The cosmic-string-only models in the two cosmologies are shown in Fig.~\ref{fig:median_GWSpectrum_ng15}(a). The kination scenario produces a modest enhancement toward the high-frequency end compared to $\Lambda$CDM although the cosmic-string tension posteriors remain largely consistent between $\mathcal{H}_1$ and $\mathcal{H}_3$ (Fig.~\ref{fig:cs_posterior_ng15}). Fig.~\ref{fig:median_GWSpectrum_ng15}(b) reveals that once contributions from SMBHBs are included in the two scenarios, the spectra become more similar, with the median spectra and the credible intervals overlapping significantly over the frequencies probed by NG15. This indicates that the additional degrees of freedom from SMBHBs can effectively absorb the spectral differences between the two cosmological frameworks and therefore the modifications associated with a non-standard cosmological expansion history can only be weakly resolved by NG15.

\begin{figure*}
    \centering
    \includegraphics[width=\textwidth]{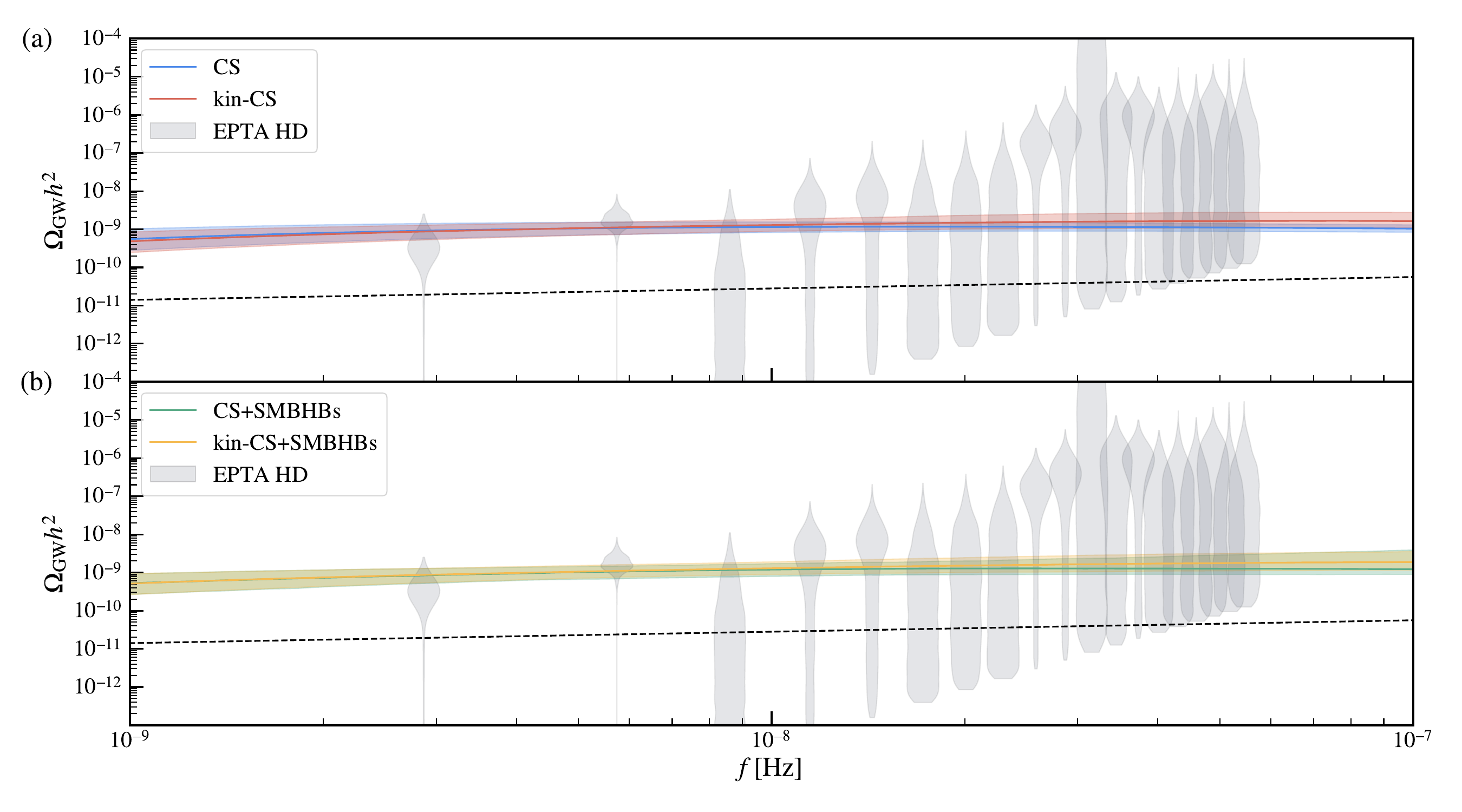}
    \caption{The median GW spectra induced by the four hypotheses for the EPTA dataset. The grey violins represent the HD-correlated free-spectrum posteriors. (a) Cosmic-string spectra in $\mathrm{\Lambda}$CDM (CS, blue) and kination (kin-CS, red) cosmologies. (b) The corresponding spectra superimposed with an additional SMBHB contribution (CS+SMBHBs, green; kin-CS+SMBHBs, orange). The shaded bands indicate the 68\% credible intervals. The black dashed lines represent the GW signal from SMBHBs with $\log_{10}A_{\rm SMBHB}=-15.6$ and $\gamma_{\rm SMBHB}=4.7$, which are the mean values from astrophysical simulations~\cite{NANOGrav:2023hvm}.}
    \label{fig:median_GWSpectrum_epta}
\end{figure*}

Figure~\ref{fig:median_GWSpectrum_epta} shows the SGWB spectra for the EPTA DR2 New+ dataset. Similar to the NG15 results, the SGWB spectra for the cosmic-string-only model in the $\Lambda$CDM and kination scenarios are broadly consistent across the frequency range probed by EPTA, with substantial overlap between their 68\% credible intervals as shown in figure~\ref{fig:median_GWSpectrum_epta}(a). While the inferred string tensions remain consistent between the two cosmologies, the presence of a non-standard cosmology leads to a slight enhancement of the reconstructed spectrum toward the high-frequency end. When an additional SMBHB contribution is included, however, the spectral differences between the two cosmological scenarios become less pronounced, with the reconstructed spectra remaining largely overlapping across the EPTA frequency range.

\section{Conclusion and outlook}\label{sec:conclusion}
Our primary objective here is to investigate how a non-standard expansion history featuring an early matter-dominated era followed by a kination phase between BBN and recombination can leave characteristic imprints on the SGWB sourced by cosmic strings in the nanohertz regime probed by NG15 and EPTA DR2 New+ pulsar timing array datasets. We perform a comprehensive Bayesian analysis and model comparison for cosmic strings in both $\Lambda$CDM and kination cosmologies, with and without additional SMBHB contributions. 

Our analysis finds no strong evidence favoring any of the alternative hypotheses over the SMBHB-only baseline $\mathcal{H}_0$, in either dataset. The underperformance of the cosmic-string model in describing NG15 arises from the preference for a blue-tilted  $\Omega_{\rm GW}h^2$ spectrum. A blue tilt can be accommodated within the cosmic-string framework by moving toward smaller string tensions, but the resulting reduction in the overall amplitude makes it difficult to account for the large amplitude inferred by the dataset. Consequently, the cosmic-string spectrum in $\Lambda$CDM is either too weak or too flat to simultaneously account for the inferred amplitude and spectral tilt. The kination scenario considered in this work enhances the SGWB spectrum toward the high-frequency end, resulting in a more pronounced blue tilt and thereby providing a better fit compared to the counterpart in $\Lambda$CDM. In contrast to NG15, the EPTA DR2 New+ data slightly favor a flatter SGWB spectrum $\Omega_{\rm GW}h^2$, which is readily accommodated by larger cosmic-string tension. This is consistent with the higher median values of $\log_{10}G\mu$ inferred from EPTA relative to NG15 across four alternative models. Nonetheless, within the cosmic string framework, with or without contributions from SMBHBs, both datasets mildly favor kination over $\Lambda$CDM, with the strength of preference ranging from negligibly small in EPTA to substantial for the cosmic-string-only model in NG15. In spite of the differences in evidence, the inferred cosmic-string-tension posteriors remain largely consistent between the two cosmologies. The $\Omega_{\rm GW}h^2$ credible intervals also show substantial overlap across the frequency range probed by NG15 and EPTA DR2 New+. We also show that the current EPTA DR2 New+ dataset does not provide sufficient sensitivity to distinguish the kination and $\Lambda$CDM expansion histories within the cosmic-string framework. The NG15 dataset shows a stronger preference for kination with the cosmic-string-only model providing substantial evidence $\ln B^{\rm NG15}_{31}=1.29\pm 0.06$. This preference diminishes once SMBHBs are included. 

A joint analysis of the NG15 and EPTA DR2 NEW+ datasets could, in principle, improve the statistical sensitivity and provide tighter constraints on model parameters and model evidences. However, the two PTA datasets contain overlapping pulsars; combining the two likelihoods will double count information associated with the common pulsars, potentially biasing the model constraints and model evidences. A statistically consistent joint analysis will therefore require dedicated statistical treatments of the correlations between overlapping datasets. The forthcoming International Pulsar Timing Array Third Data Release (IPTA DR3)~\cite{Shaifullah:2026xcq} will combine observations from multiple PTA collaborations and will provide a more comprehensive joint dataset for characterizing the SGWB. This is particularly relevant for the present work since NG15 and EPTA DR2 New+ exhibit different spectral preferences and yield somewhat different statistical supports for kination cosmology. A consistent joint analysis will clarify the nature of the SGWB and its implication for the early expansion history, which we leave to future work.

\section*{Acknowledgments}
We acknowledge Kai Schmitz and Sai Chaitanya Tadepalli for helpful discussions. We thank Akshay Ghalsasi for his help on \texttt{PRyMordial}. This work was supported by the Department of Energy under Grant No.~DE-SC0025611. For facilitating portions of this research, R.C.~wishes to acknowledge the Center for Theoretical Underground Physics and Related Areas (CETUP*), the Institute for Underground Science at Sanford Underground Research Facility (SURF), and the South Dakota Science and Technology Authority for hospitality and financial support, as well as for providing a stimulating environment when this work was finalized. This research was supported in part by Lilly Endowment, Inc., through its support for the Indiana University Pervasive Technology Institute.

\bibliographystyle{JHEP.bst}
\bibliography{ref} 
\end{document}